\documentclass[aps,12pt,amsmath,showpacs]{revtex4}
\usepackage{graphicx}
\usepackage{subfigure}
\usepackage{amssymb}
\begin{document}
\title{Gain/loss induced localization in one-dimensional PT-symmetric tight-binding models}
\author{O. V\'azquez-Candanedo$^1$, J. C. Hern\'{a}ndez-Herrej\'{o}n$^2$,
F. M. Izrailev$^{1,3}$, D. N. Christodoulides$^4$}
\address{$^1$Instituto de F{\'\i}sica, Universidad Aut\'onoma de Puebla, Apartado Postal J-48,
Puebla, 72570 Mexico}
\affiliation{$^2$Instituto de F{\'\i}sica, Universidad Aut\'onoma de San Luis Potos\'i, Av. Manuel Nava 6, Zona Universitaria, 78290 Mexico}
\affiliation{$^3$NSCL and Dept. of Physics and
Astronomy, Michigan State University - East Lansing, Michigan
48824-1321, USA}
\affiliation{$^4$CREOL/College of Optics, University of Central Florida, Orlando, Florida 32816, USA}
\date{\today}
\begin{abstract}
We investigate the properties of PT-symmetric tight-binding models by considering both bounded and unbounded models. For the bounded case, we obtain closed form expressions for the corresponding energy spectra and we analyze the structure of eigenstates as well as their dependence on the gain/loss contrast parameter.  For unbounded PT-lattices, we explore their scattering properties through the development of analytical models. Based on our approach we identify a mechanism that is responsible to the emergence of localized states that are entirely due to the presence of gain and loss. The derived expressions for the transmission and reflection coefficients allow one to better understand the role of PT-symmetry in energy transport problems occurring in such PT-symmetric tight-binding settings. Our analytical results are further exemplified via pertinent examples.
\end{abstract}

\maketitle
\section{Introduction}
Over the years the transport properties of Hermitian lattice systems have been a subject of intense investigation. Such arrangements are ubiquitous in nature and are typically characterized by succession of allowed bands and forbidden gaps \cite{1}. They appear in many and diverse fields of physics and applied physics ranging from solid state, to Bose-Einstein condensates \cite{2}, to optical photonic crystals and lattices \cite{3,4}. On the other hand, much less attention has been paid to non-conservative periodic structures. In this case, a configuration can in general exhibit either gain or loss and hence is non-Hermitian. While quantum mechanics is by nature Hermitian, gain/loss can be readily incorporated in classical settings.
	In the late nineties, the notion of parity-time symmetry was first introduced by Bender and Boettcher within the framework of quantum field theory \cite{5}. In this work it was shown that a broad family of Hamiltonians can exhibit entirely real spectra as long as they commute with the parity-time (PT) operator and hence they may share a common set of eigenvectors. This is to some extent unexpected given that such properties are typically associated with Hermitian systems. In general, the eigen-energies of such arrangements are real only within a certain range of a non-hermiticity parameter. Yet, once this parameter exceeds a critical threshold, the system can undergo a spontaneous symmetry breaking, corresponding to a transition from real to complex spectra thus entering the so-called broken PT-symmetry regime \cite{5,6}. Interestingly, this phase transition point exhibits all the characteristics of an exceptional point singularity.

In recent years the possibility of observing PT-symmetric effects in optics has been suggested \cite{7,8,9,10}. In this context PT symmetry can be readily established by requiring that the complex refractive index distribution obeys the relation $n(x) = n^{\star}(-x)$. In other words, this symmetry demands that the refractive index profile must be an even function of position while the gain/loss spatial distribution should be antisymmetric. As pointed out in several studies \cite{11,12,13,14,15,16,17,18,19,20,21,22,23,24,25,26,27,28,29}, PT-symmetry can lead to a number of intriguing processes. These include for example band-merging effects in PT-symmetric lattices \cite{8}, abrupt phase transitions \cite{12}, power oscillations and double refraction, and unidirectional invisibility \cite{14,16,24}. In addition, non-reciprocal wave propagation is also possible when PT-symmetry is used in conjunction with nonlinearity \cite{19}. Other issues like defect states in PT-lattices \cite{14}, the coexistence of coherent lasing-absorbing modes \cite{17,23}, and mode selection in PT-symmetric lasers have also been investigated in the literature \cite{25}.

In this paper we present the results of a detailed study of the one-dimensional PT-symmetric tight-binding model consisting of alternating gain/loss sites. Our main interest is in establishing the relation between the properties of the isolated model with finite number of sites with those of scattering in the case when the same structure is attached to perfect leads. For the isolated model we are interested in the energy spectrum and structure of eigenstates emerging due to fixed boundary conditions. Contrary, for the scattering problem the question is about the scattering states and global characteristics of transmission and reflection in dependence on the model parameters. So far, in literature the main results concern either the energy spectrum of the bounded model, or the band structure of energy for transmitted waves. Both the relation between bounded and unbounded models, as well as the structure of eigenstates and scattering states is not considered properly.

Studying the structure of eigenstates, we show that even without the presence of disorder one can speak about the localization defined via the exponential decrease of the transmission coefficient. Although the physical effect of such a localization in the presence of a gain only (without absorption) is already studied, the mechanism of this unexpected effect was not fully understood. In our approach we show that in PT-symmetric models a similar effect also emerges, however, it is much more complicated due to the interplay between gain and loss. By studying the structure of scattering states we have found that they are, indeed, exponentially localized under some conditions. In this case, the localization length of the scattering states (and not of the eigenstates of the isolated model that remain to be extended) is the same as that defined by the decrease of the transmission coefficient. We were able to derive the analytical expression for the transmission coefficient, obtained for any values of the control parameters. This expression explains the properties of the transmission depending on the energy of scattering states, strength of gain/loss and length of the scattering structure.

\section{The Model}

We consider the one-dimensional tight-binding model which is
described by the standard Hamiltonian,
\begin{equation}
 H_{mn}=\epsilon_n\delta_{mn}+\nu(\delta_{n,n+1}+\delta_{n,n-1}),
 \label{hpt}
\end{equation}
where $\nu$ is the hopping amplitude connecting the nearest sites
(in what follows we fix $\nu=1$). As for imaginary on-site potential
$\epsilon_n$, its form is defined as follows,
\begin{equation}
  \epsilon_n=\left\{ \begin{array}{ll}
-i\gamma&\quad\text{for $n$ odd},\\
i\gamma &\quad\text{for $n$ even},
\end{array}\right.
 \label{pot_pt}
\end{equation}
where $\gamma>0$ stands for the loss (for $n$ even) or for the gain (for $n$ odd). This model can be treated as the bi-layer model with alternating gain/loss sites, thus creating the structure belonging to the class of PT-symmetric models revealing quite unexpected properties of scattering {see, for example, \cite{13,15,18,24,BFKS09,Ro12}).

The Schr\"{o}dinger equation with non-Hermitian Hamiltonian (\ref{hpt}) takes the form,
\begin{equation}
i \hbar \frac{d\Psi_n(t)}{dt} = \Psi_{n+1}(t)+\Psi_{n-1}(t)+\epsilon_{n}\Psi_{n}(t).
\label{1}
\end{equation}
The solution of this equation can be presented in the conventional form,
\begin{equation}
 \Psi_n(t)=e^{-iEt}\psi_n,
\label{1a}
\end{equation}
with $E$ as the energy of an eigenstate $\psi_n$. As will be shown, the energy $E$ can be either real or complex (in fact, imaginary) depending on the value of $\gamma$. For non-Hermitian matrices there are two sets of eigenstates, left and right, however, we consider right eigenstates only. The relation between the two sets of eigenstates will be discussed below. Note also that there is a special case of $E=0$ which we analyze separately.

Thus, we arrive at the stationary discrete Schr\"{o}dinger equation,
\begin{equation}
E\psi_n= \psi_{n+1}+\psi_{n-1}+\epsilon_{n}\psi_{n}.
\label{1a}
\end{equation}
The general solution $\psi_n$ of this equation can be written in the form,
\begin{equation}
\psi_{n}=\left\{ \begin{array}{lll}
\delta(Ae^{ink}+Be^{-ink})&\quad\text{for $n$ odd},\\
Ae^{ink}+Be^{-ink}&\quad\text{for $n$ even},
\end{array}\right.
\label{psi-m}
\end{equation}
where  relations between $A, B$ are defined by either boundary
conditions at $n=0, 2N+1$ for the {\it bounded model} or by the
conditions at $n=-1, 0$ for {\it unbounded model}. Here and below by
bounded model we mean that apart from the gain/loss at the sites
$n$, there is no coupling to continuum at the edges of a structure.
In other words, this model corresponds to the problem of the
dynamics of wave packets in the presence of fixed or periodic
boundary conditions. Contrary, the unbounded model corresponds to the
scattering problem for which the structure of size $2N$ is attached
to perfect leads, and the main interest is in the
transmission/reflection coefficient. For both models the two
parameters $\delta$ and $k$ can be expressed in terms of energy $E$
and control parameter $\gamma$ as follows,
\begin{equation}
\begin{array}{ccc}
 E+i\gamma&=&(2/\delta)\, \cos k,\\
 E-i\gamma&=&2 \delta \, \cos k.
 \label{E-k}
\end{array}
\end{equation}

\section{Bounded model}
\subsection{Spectrum}
Considering a system where for the sites $n=1,\ldots,2N$ the potential obeys Eq.(\ref{pot_pt}) and taking zero boundary
conditions at sites $n=0$ and $n=2N+1$, i.e. $\psi_{0}=\psi_{2N+1}=0 $, we have the relations,
\begin{equation}
\begin{array}{ccc}
 A+B&=&0,\\
 Ae^{i(2N+1)k}+Be^{-i(2N+1)k}&=&0,
 \label{AB}
\end{array}
\end{equation}
which make the constants $A$ and $B$ linearly dependent and define discrete values for the parameter $k$,
\begin{equation}
 k_s=\frac{s\pi}{2N+1}
 \label{k}
\end{equation}
with $s=1,...,N$. Inserting $k_s$ into Eqs.(\ref{E-k}) one can find
the energy spectrum $E_s$ which is defined by the relation,
\begin{equation}
4\cos^2k_s=E_s^2+\gamma^2. \label{main-1}
\end{equation}
To continue, it is useful to introduce the parameter $\beta_s$,
\begin{equation}
 E_s=\pm 2 \cos k_{s}\, \sin \beta_s,
\label{spectrum}
\end{equation}
which can be expressed via $k_s$ and $\gamma$,
\begin{equation}
\cos \beta_s = \frac{\gamma}{2 \cos k_{s}}.
 \label{beta-n}
\end{equation}
According to Eqs.(\ref{E-k}) $\delta$ can be written as,
\begin{equation}
 \delta_{\pm}^{(s)} =-i e^{\pm i\beta_s}.
 \label{2-delta}
\end{equation}
The plus/minus signs in Eq.(\ref{spectrum}) stand to stress that for
any value of $k_s$ there are two values of energy symmetric with
respect to the band center $E=0$. From Eq.(\ref{beta-n}) one can see
that $\beta_s$ can take real or imaginary values depending on whether
$\gamma$ is smaller or larger than $2 \cos k_{s}$, respectively.
Therefore, the energy $E_s$ in Eq.(\ref{spectrum}) can be either
real or purely imaginary, the result which is entirely due to the
PT-symmetry of our gain/loss potential.

Let us now analyze the properties of the energy spectrum in
dependence on the parameter $\gamma$ for a fixed $N$. From
Eq.(\ref{beta-n}) one gets that all eigenvalues are real for $\gamma
< 2 \cos k_s$ for any value of $s$. Since the smallest value of
$\cos k_s$ occurs for $s=N$, the condition of a completely real
spectrum is
\begin{equation}
\gamma < \gamma_{cr}^{(1)} = 2 \cos \left(\frac{N\pi}{2N+1}\right) \approx \frac{\pi}{2N},
\label{gamma-1}
\end{equation}
where the estimate for $N \gg 1$ is also given. A typical example of
such spectrum is shown in Fig.\ref{interm}a for $\gamma=0.05$ and
$N=10$, for which $\gamma_{cr}^{(1)} = 0.157$. Note that in this
case imaginary parts of $E_s$ vanish.

On the other hand, when
\begin{equation}
\gamma > \gamma_{cr}^{(2)} = 2 \cos \left(\frac{\pi}{2N+1}\right)
\approx 2 \left(1-\frac{\pi^2}{8N^2}\right), \label{gamma-2}
\end{equation}
all values of $E_s$ are imaginary. Therefore, for $
\gamma_{cr}^{(1)}< \gamma < \gamma_{cr}^{(2)}$ some of the
eigenvalues $E_s$  are real and others are imaginary, see
Fig.\ref{interm}a. The data in this figure demonstrate that all
eigenvalues $E_s$ are combined in pairs, having the symmetry (with
respect to zero) either for real or pure imaginary values.
\begin{figure}[ht]
\centering \subfigure[]{\includegraphics[scale=0.37]{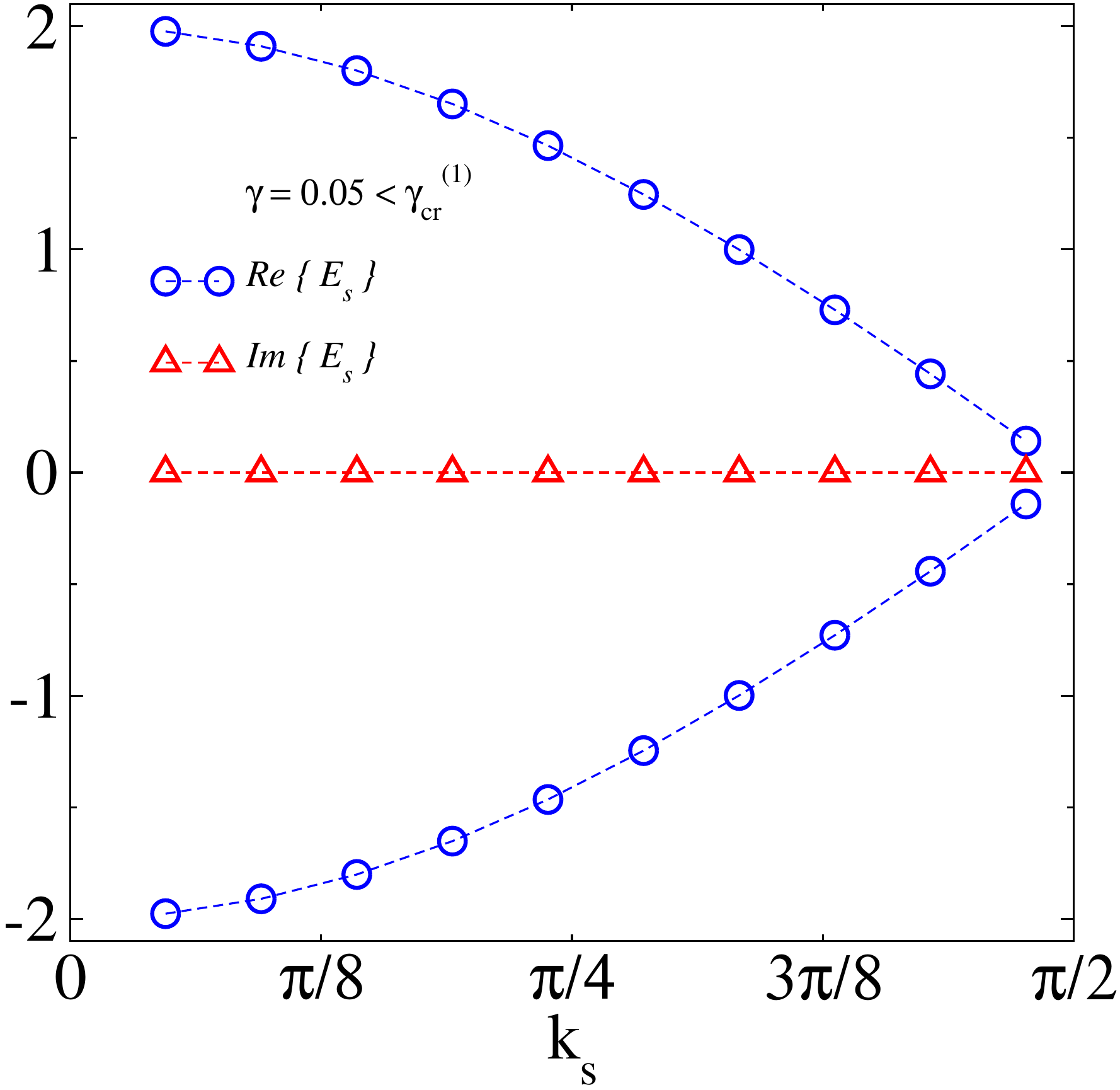}}
\subfigure[]{\includegraphics[scale=0.37]{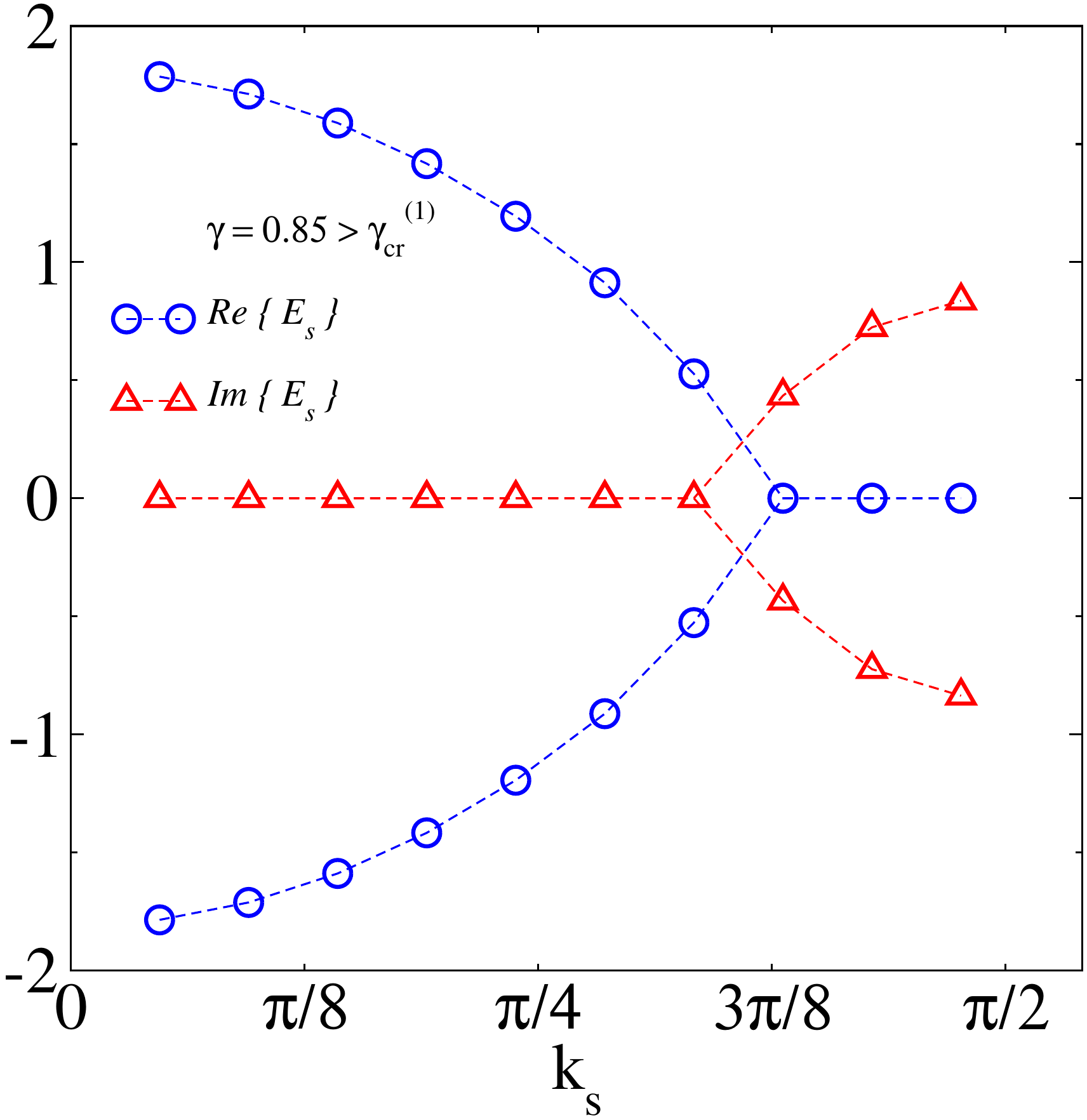}}
\subfigure[]{\includegraphics[scale=0.37]{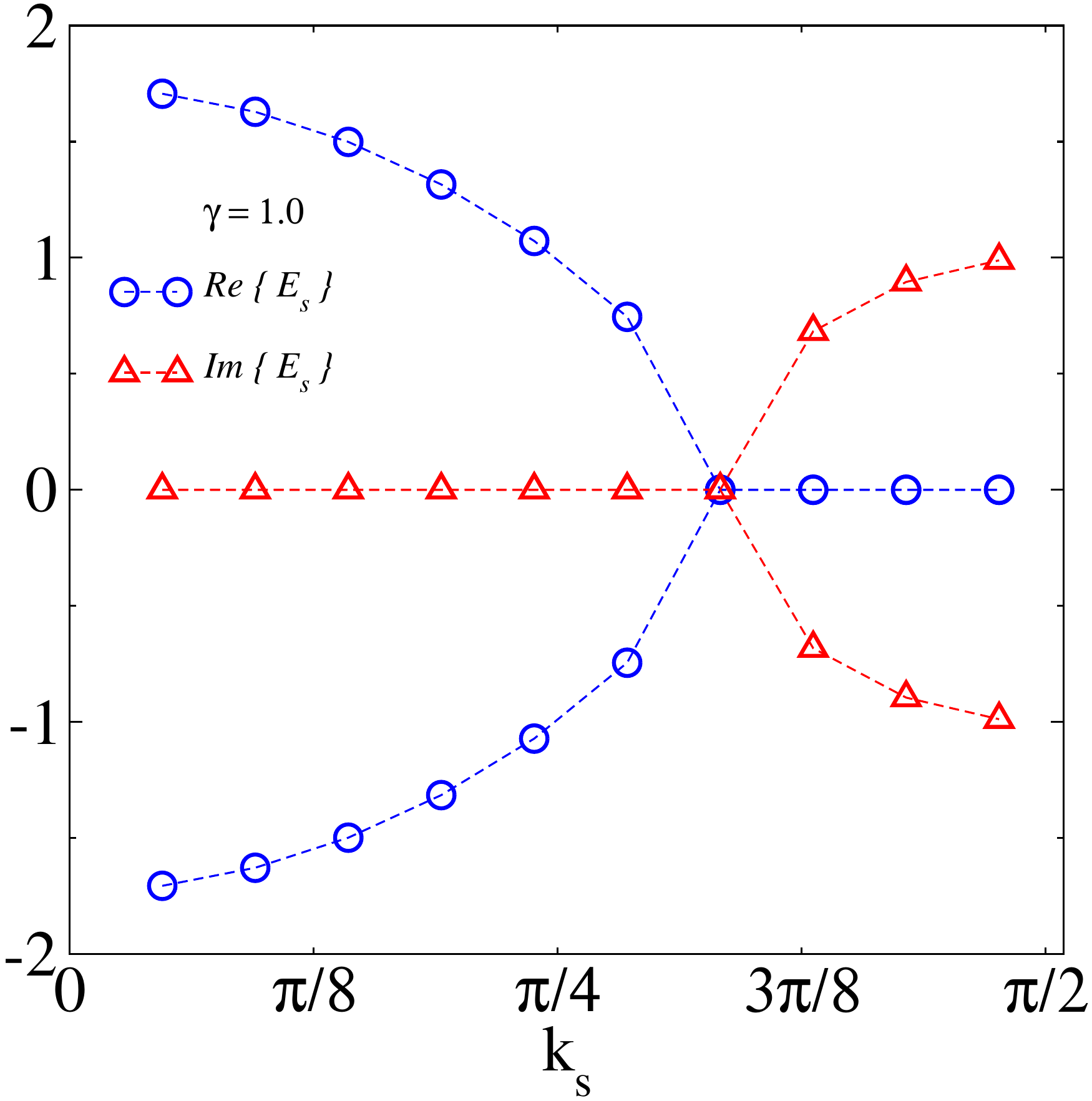}}
\subfigure[]{\includegraphics[scale=0.37]{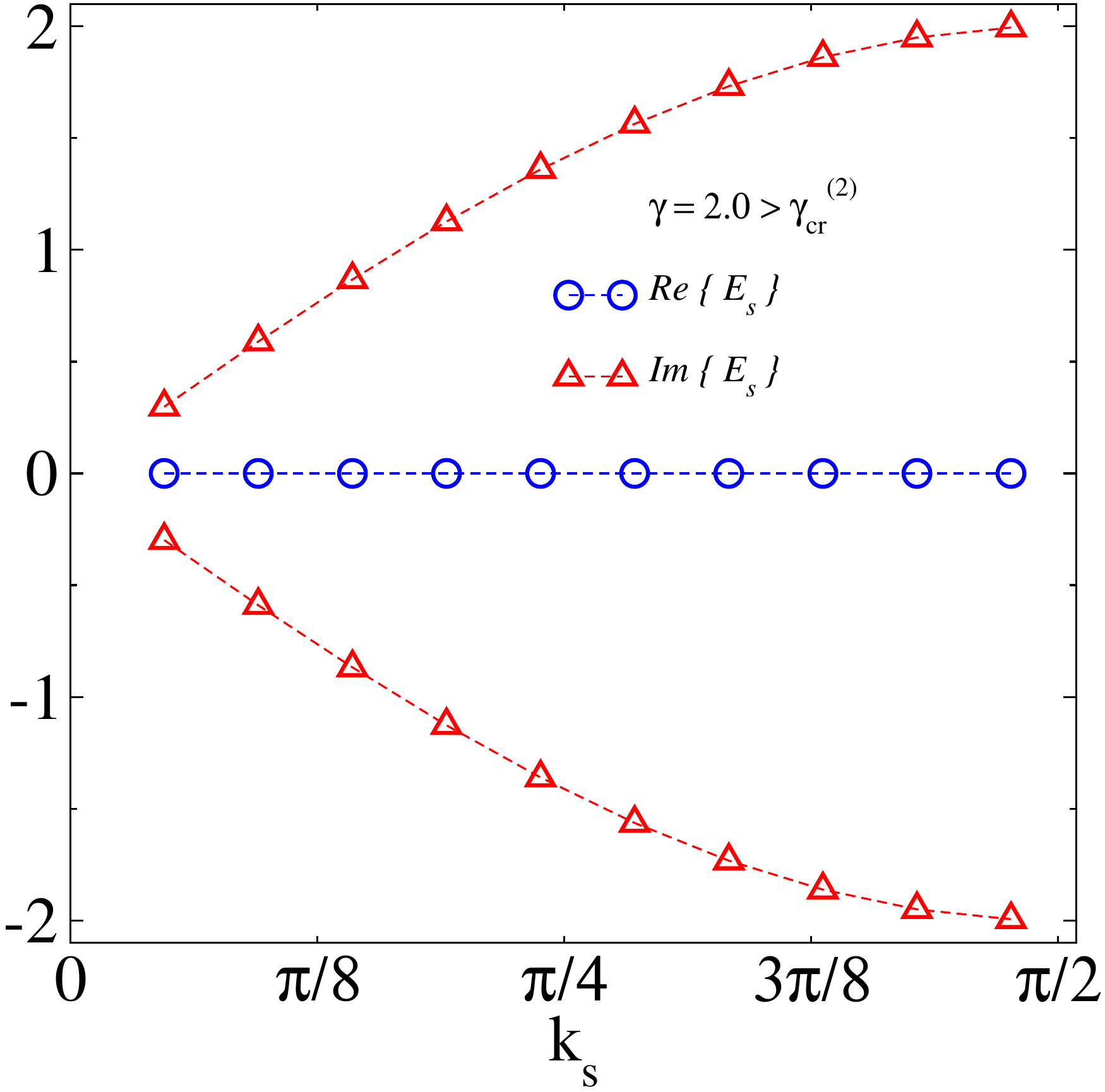}}
\caption{
\footnotesize (Color online) Energy spectrum $E_s$ as a function of
$k_s$ for $N=10$: (a) $\gamma=0.05$ with $\gamma<\gamma_{cr}^{(1)}$,
therefore, all eigenvalues are real; (b) $\gamma=0.85$ with
$\gamma>\gamma_{cr}^{(1)}$ for which there are both real and
imaginary eigenvalues; (c) $\gamma=\gamma_{cr}=1.0$ for $s=N-3$, such
that two eigenvalues coalesce at the band center; (d) $\gamma=2.0$
with $\gamma>\gamma_{cr}^{(2)}$ when all eigenvalues are imaginary.}
\label{interm}
\end{figure}
\begin{figure}[ht]
\centering \subfigure[]{\includegraphics[scale=0.37]{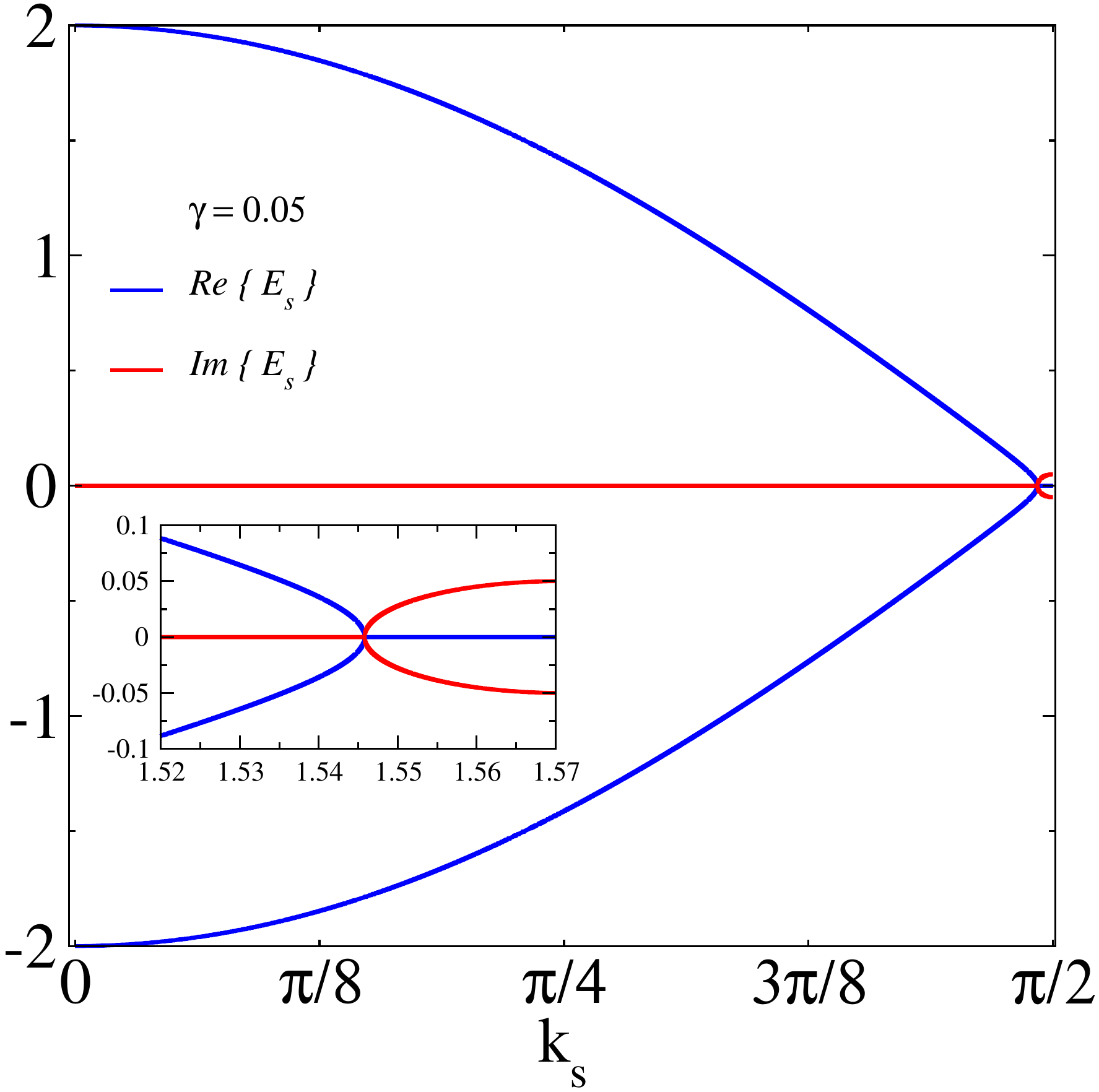}}
\subfigure[]{\includegraphics[scale=0.37]{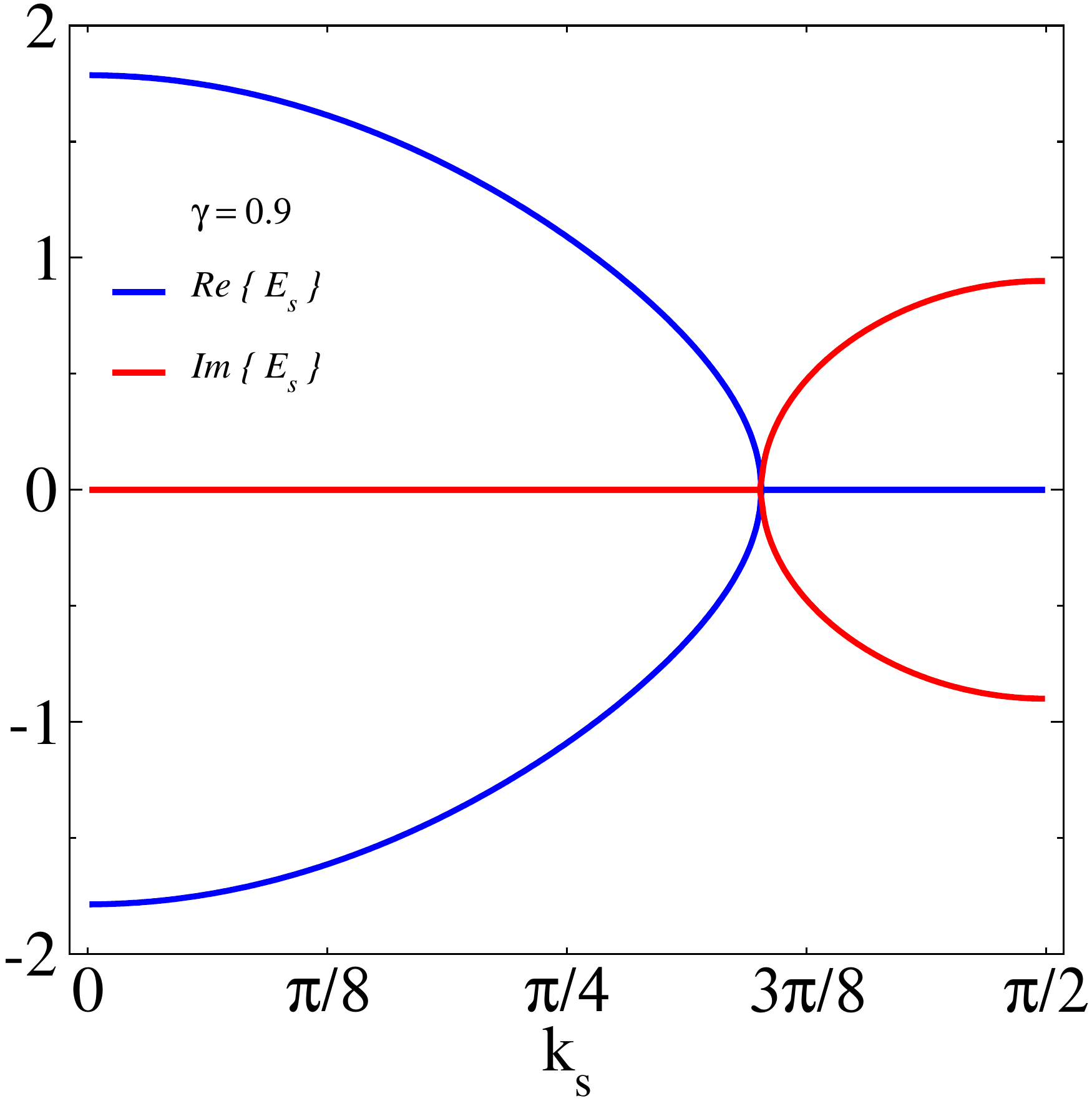}} \caption{
\footnotesize (Color online) Energy spectrum for system with $N
\rightarrow \infty$: (a) Spectrum for $\gamma=0.05$, inset shows zoom of the region around
critical value; (b) Spectrum for $\gamma=0.9$. }
\label{interm-1}
\end{figure}
From Fig.\ref{interm} one can understand the dynamics of energy
levels upon the increase of $\gamma$. For $\gamma=0$ all values of
$E_s$ are real in the range $-2 < E_s < 2$ and correspond to those
emerging in the perfect lattice. With an increase of
$\gamma$, two eigenvalues mostly close to the band center start to
move to zero, and after they continue to move along the imaginary
axes, one moves up and another moves down. The critical point
$\gamma = \gamma_{cr}^{(1)}$ is known in literature as an {\it
exceptional point} at which both real and imaginary parts of $E_s$
vanish.

With further increase of $\gamma$ the second pair of real
eigenvalues closest to $E=0$ approaches
the band center. When passing the corresponding critical value
(second exceptional point) for this pair of $E_s$, one eigenvalue
goes up along the imaginary axes, and another goes down. This
scenario continues with an increase of $\gamma$, and when $\gamma >
\gamma_{cr}^{(2)}$ all the values $E_s$ are purely imaginary. It is
clear that in the limit $N \rightarrow \infty$, the critical value
$\gamma_{cr}^{(1)}$ vanishes, thus indicating that the phase in
which all eigenvalues are real, is absent, see Fig.\ref{interm-1}a,b.

\subsection{Eigenstates}

For non-Hermitian Hamiltonians with a discrete spectrum there are
two sets of eigenstates, left ($\psi_l^{(s)}$) and right
($\psi_r^{(s)}$) defined by the equations,
\begin{equation}
 H\psi_r^{(s)}=E_s\psi_r^{(s)},\quad \psi_l^{(s)}H=E^s\psi_l^{(s)},
 \label{lr-se}
\end{equation}
with, in general, complex conjugate eigenvalues, $E^s=E_s^*$,
\cite{W03}. Since, $\psi_l^{(s)}=(\psi_r^{(s)})^*$ we explore
one set of eigenstates only. In what follows we solve the first of
Eqs.(\ref{lr-se}) only, determining the structure of right
eigenstates. For this reason we omit the index $r$
for right eigenstates.

To start with the global structure of eigenstates in connection with
the properties of spectra, first, one has to understand the
symmetric properties of eigenstates. Because the parameter $\delta$
in Eq.(\ref{2-delta}) takes two values, there are two types of
eigenstates, $\psi_n^+$ and $\psi_n^-$. The right eigenstates can be
obtained by applying the conditions obtained from Eqs.(\ref{AB}) to
Eq.(\ref{psi-m}), which gives,
\begin{equation}
 \psi_n^{(s)\pm}=\left\{ \begin{array}{ll}
2i\delta_{\pm}^{(s)}A_{\pm}^{(s)}\sin(nk_s)&\quad\text{for $n$ odd},\\
2iA_{\pm}^{(s)}\sin(nk_s)&\quad\text{for $n$ even}.
\end{array}\right.
\label{psi-gen}
\end{equation}
Here $A_{\pm}^{(s)}$ is a constant that can be determined by the normalization condition.

\subsubsection{Eigenstates for $\gamma<2\cos k_s $}

For $\gamma<2\cos k_s$ the parameter $\beta_s$ is real, see
Eq.(\ref{beta-n}). Normalizing the eigenstates such that
$\sum_{n=1}^{2N}|\psi_{n}^{(s)\pm}|^2=1$ one gets,
\begin{equation}
 |A_{+}^{(s)}|^2=|A_{-}^{(s)}|^2=\frac{1}{\zeta^2}
 \label{pt-A}
\end{equation}
where
\begin{equation}
 \zeta=\sqrt{2(2N+1)}.
 \label{pt-zeta}
 \end{equation}
Note that the factors $A_{\pm}^{(s)}$ are determined up to some
phase which we chose in such a way that for $\gamma=0$ the standard
expressions for eigenstates in the perfect lattice are recovered.
Therefore, by introducing the following relations,
\begin{equation}
A_+^{(s)}=\frac{e^{-i\beta_s}}{\zeta},\quad \nu_s=\frac{\pi}{2}-\beta_s, \quad
\quad\mathcal{D}=\frac{2}{\zeta}, \label{constants-1}
\end{equation}
the states $\psi_n^{(s)+}$ for positive $E_s>0$ get the form,
\begin{equation}
\psi_n^{(s)+}=\left\{ \begin{array}{ll}
\mathcal{D}\sin(nk_s)&\quad\text{for $n$ odd},\\
e^{i\nu_s}\mathcal{D}\sin(nk_s)&\quad\text{for $n$ even}.
\end{array}\right.
\label{psi+}
\end{equation}
In the same way, by introducing
\begin{equation}
A_-^{(s)}=\frac{i}{\zeta}, \label{constants-2}
\end{equation}
the other set $\psi_n^{(s)-}$ of eigenstates with negative energies,
$E_s < 0$, is defined by
\begin{equation}
\psi_n^{(s)-}=\left\{ \begin{array}{ll}
e^{i\nu_{s}}\mathcal{D}\sin(nk_s)&\quad\text{for $n$ odd},\\
-\mathcal{D}\sin(nk_s)&\quad\text{for $n$ even}.
\end{array}\right.
\label{psi-}
\end{equation}

An example of eigenstates for energies $E_{s}=\pm1.46$ is shown in
Fig.\ref{eigenstates} for a system with $\gamma=0.05$ and $N=10$.
The most important issue of such eigenstates with real energies is
that all of them are extended in the site representation. This fact
is due to the fixed boundary conditions, as it also happens in
classical chains of linear oscillators (similar structure occurs for
periodic boundary conditions). As for the time-dependent part of the
solution $\Psi_n(t)$ (see Eq.(\ref{1a})), since the energies $E_s$
are real, each site oscillates with the same frequency.

The detailed analysis of the eigenstates shows quite interesting
symmetries between $\psi_n^{(s)+}$ and $\psi_n^{(s)+}$.
Specifically, for $n$ odd
\begin{figure}[ht]
\centering \subfigure[]{\includegraphics[scale=0.3]{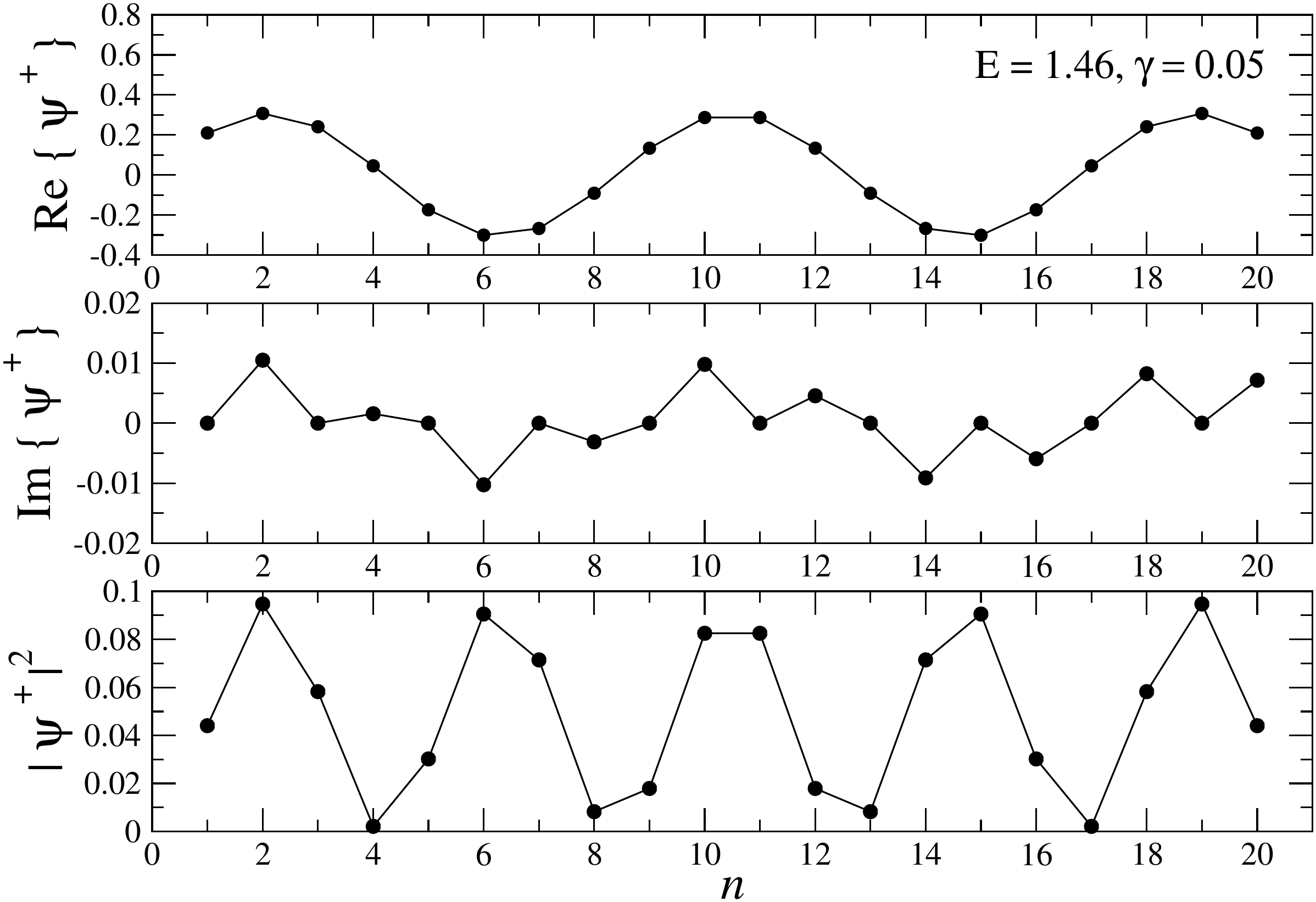}}
\subfigure[]{\includegraphics[scale=0.3]{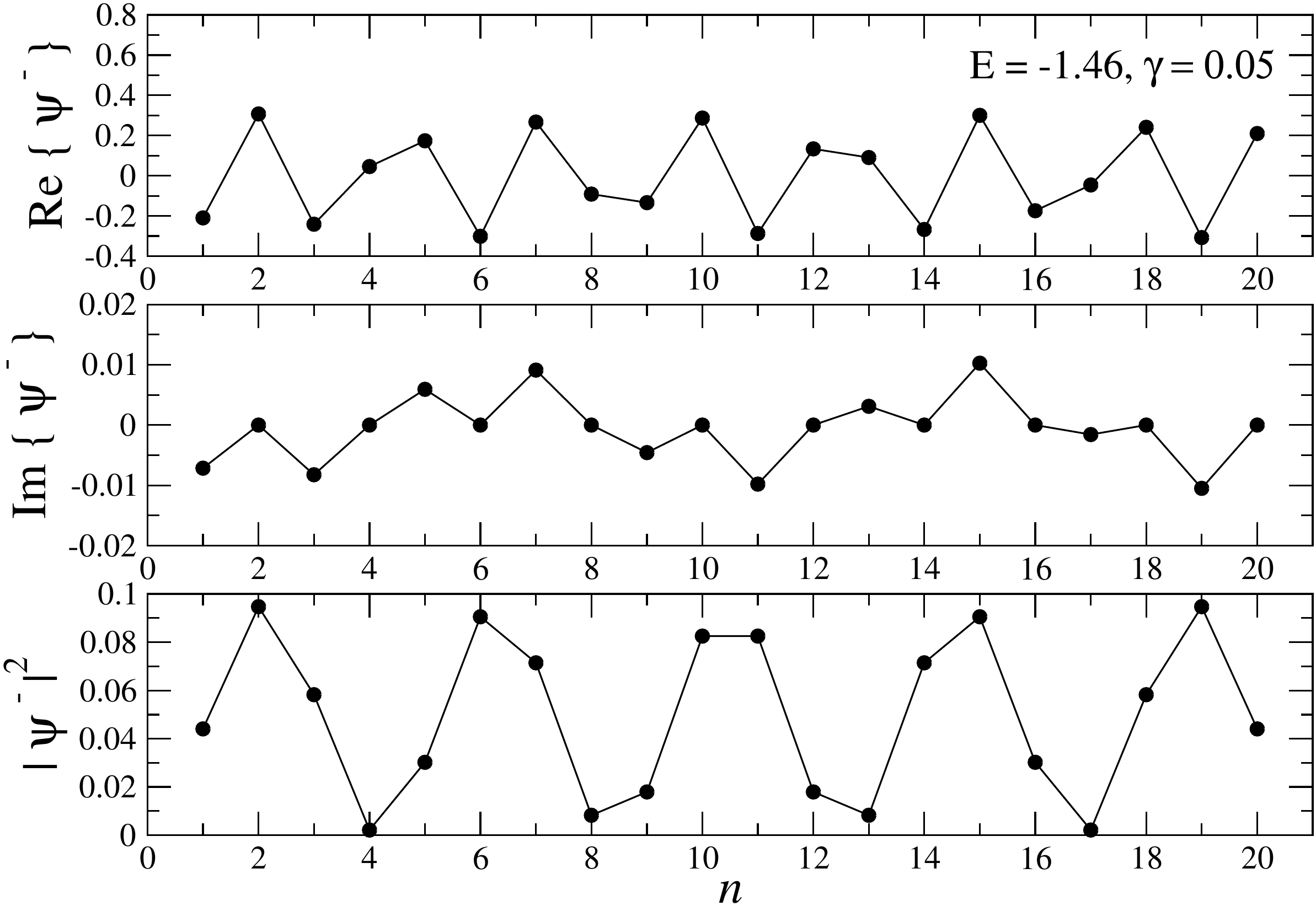}} \caption{ The
eigenstates corresponding to the energy $E_{s}=1.46$ (left panel),
and to $E_{s}=-1.46$ (right panel) for a system with $\gamma=0.05$
and $N=10$.} \label{eigenstates}
\end{figure}
one can reveal the relations,
\begin{equation}
\psi_n^{(s)+}=\left\{ \begin{array}{lll}
(\psi_{2N-n+1}^{(s)-})^*,&for\quad s\quad odd,\\
-(\psi_{2N-n+1}^{(s)-})^*,&for\quad s\quad even.
\end{array}\right.
\end{equation}
Correspondingly, for $n$ even one gets,
\begin{equation}
\psi_n^{(s)+}=\left\{ \begin{array}{lll}
-Re\{(\psi_{2N-n+1}^{(s)-})^*\}+Im\{(\psi_{2N-n+1}^{(s)-})^*\},&for\quad s\quad odd,\\
Re\{(\psi_{2N-n+1}^{(s)-})^*\}-Im\{(\psi_{2N-n+1}^{(s)-})^*\},&for\quad
s\quad even.
\end{array}\right.
\end{equation}
These symmetries can be seen in Fig.\ref{eigenstates} under a close
inspection.

\subsubsection{Eigenstates for $\gamma>2\cos k_s$}

According to Eq.(\ref{beta-n}) in this case $\beta_s$ is imaginary,
therefore, $\beta_s=i\tilde{\beta_s}$ with $\tilde\beta_s$ real. The
normalization will be now given by
\begin{equation}
|A_{\pm}^{(s)}|^2=\frac{1}{\zeta_{\pm}^{(s)2}},
 \label{pt-An}
\end{equation}
where
\begin{equation}
 \zeta_\pm^{(s)}=\sqrt{(e^{\mp2\tilde{\beta_s}}+1)(2N+1)}.
 \label{pt-zetan}
 \end{equation}

By introducing
 \begin{equation}
A_+^{(s)}=\frac{-i}{\zeta_+^{(s)}}\quad
\text{and}\quad\mathcal{D_+}=\frac{2}{\zeta_+^{(s)}},
\label{constants-3}
\end{equation}
the eigenstate $\psi_n^{(s)+}$ for positive imaginary part $Im
{\{E_s\}} >0$ can be presented in the form,
\begin{equation}
\psi_n^{(s)+}=\left\{ \begin{array}{lll}
i\mathcal{D_+}e^{-\beta_s}\sin(nk_s)&\quad\text{for $n$ odd},\\
\mathcal{D_+}\sin(nk_s)&\quad\text{for $n$ even}.
\end{array}\right.
\label{psi+n}
\end{equation}
Also, by using
\begin{equation}
A_-^{(s)}=-\frac{1}{\zeta_-^{(s)}}\quad
\text{and}\quad\mathcal{D_-}=\frac{2}{\zeta_-^{(s)}},
\label{constants-4}
\end{equation}
the other set $\psi_n^{(s)-}$ of eigenstates with negative imaginary
part, $Im {\{E_s\}} < 0$, takes the form,
\begin{equation}
\psi_n^{(s)-}=\left\{ \begin{array}{lll}
\mathcal{D_-}e^{\beta_s}\sin(nk_s)&\quad\text{for $n$ odd},\\
-i\mathcal{D_-}\sin(nk_s)&\quad\text{for $n$ even}.
\end{array}\right.
\label{psi-n}
\end{equation}

Fig.\ref{eigen-n} demonstrates the structure of two eigenstates
corresponding to energy $E_s=\pm i0.133$, for a system with
$\gamma>2\cos k_N$ and $N=10$. In this case there are only two
eigenstates with imaginary energies. One can see that according to
Eqs.(\ref{psi+n},\ref{psi-n}) both eigenstates are extended in the position
representation. However, in contrast with the previous case of real
energies $E_s$ now the modes $\Psi_n(t)$ are either exponentially
increasing or decreasing in time, depending on the sign of the imaginary
part of eigenvalues.
\begin{figure}[ht]
\centering \subfigure[]{\includegraphics[scale=0.3]{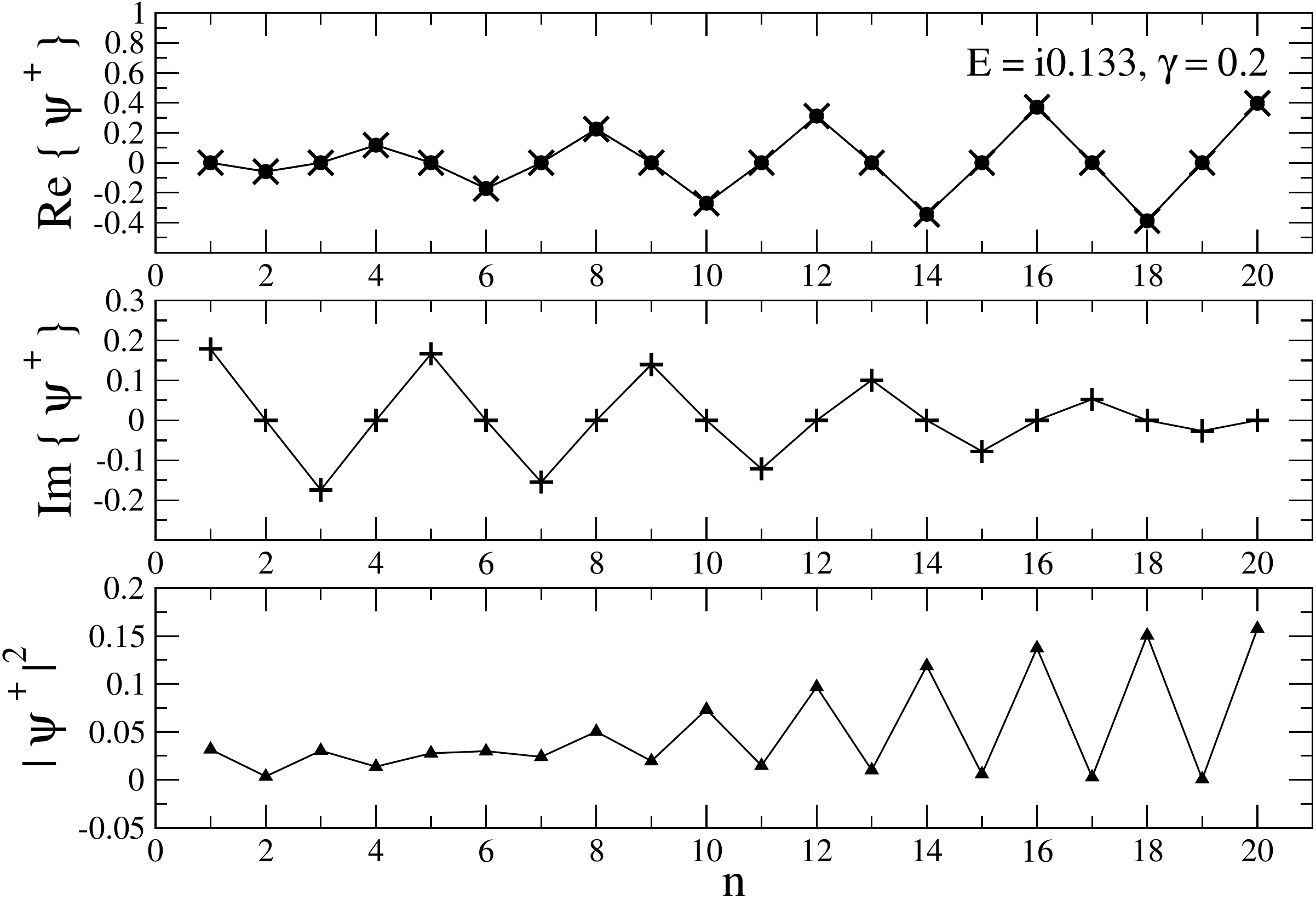}}
\subfigure[]{\includegraphics[scale=0.3]{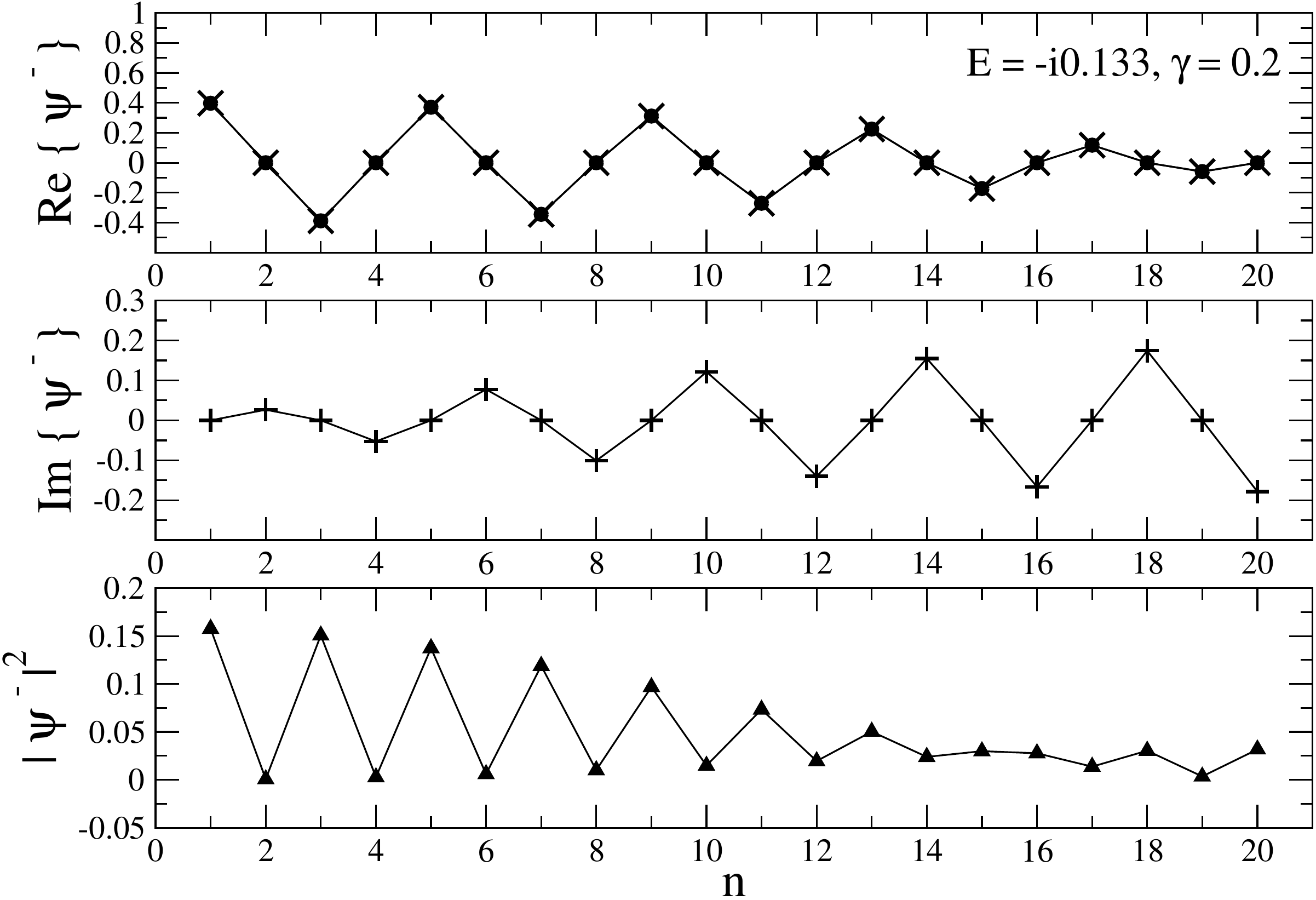}} \caption{
\footnotesize Eigenstates corresponding to energies $E_{s}=\pm
i0.133$, for a system with $\gamma>2\cos k_N$, $N=10$ and
$\gamma=0.2$.} \label{eigen-n}
\end{figure}

In Fig.\ref{eigen-n} the symmetry between $\psi_m^{(s)+}$ and
$\psi_m^{(s)-}$ is quite simple, this is due to the fact that for
the chosen parameters, $\tilde{\beta_s}\approx0$ and $\mathcal{D_+}\approx\mathcal{D_-}$.
However, as $\gamma$ grows this symmetry disappears.

\subsubsection{Eigenstates for $\gamma=2\cos k_s $}

For the case when $\gamma=2 \cos k_s$ two eigenvalues approach the
band center, $E=0$, from the left and right and the value of
$\beta_s$ vanishes. At this exceptional point the corresponding
eigenstates are related to each other due to the simple relation,
$\psi_n^{(s)+}=i \psi_n^{(s)-}$, see Eqs.(\ref{psi+}) and
(\ref{psi-}). The example of such eigenstates is given in
Fig.\ref{eigens-0} for a system with $\gamma=\gamma_{cr}^{(1)}$ and
$N=10$. As one can see, for this specific case the corresponding solution of
the time-dependent equation (\ref{1}) does not depend on time,
$\Psi_n(t)\sim \psi_n^{(s)\pm} $. However, this is not the only
solution of Eq.(\ref{1}), there is an additional solution which
linearly depends on time, $\Psi_n(t)\sim t \psi_n^{(s)\pm} $. This
fact can be easily confirmed by the direct evaluation of
Eq.(\ref{1}). The consequence of this result for the wave packet
dynamics has been discussed in Ref.\cite{24}.
\begin{figure}[ht]
\vspace{2cm}
\centering \subfigure[]{\includegraphics[scale=0.3]{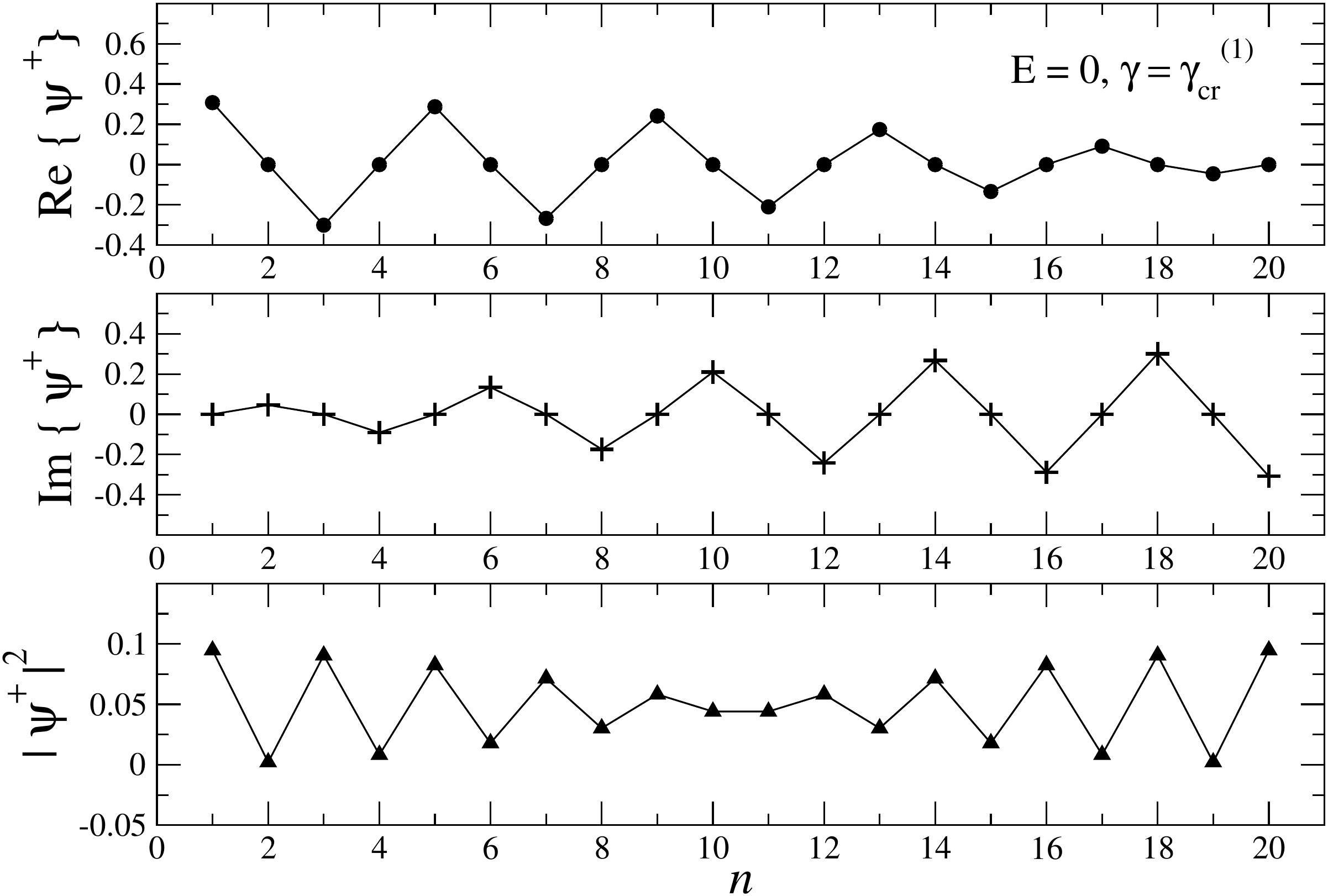}}
\subfigure[]{\includegraphics[scale=0.3]{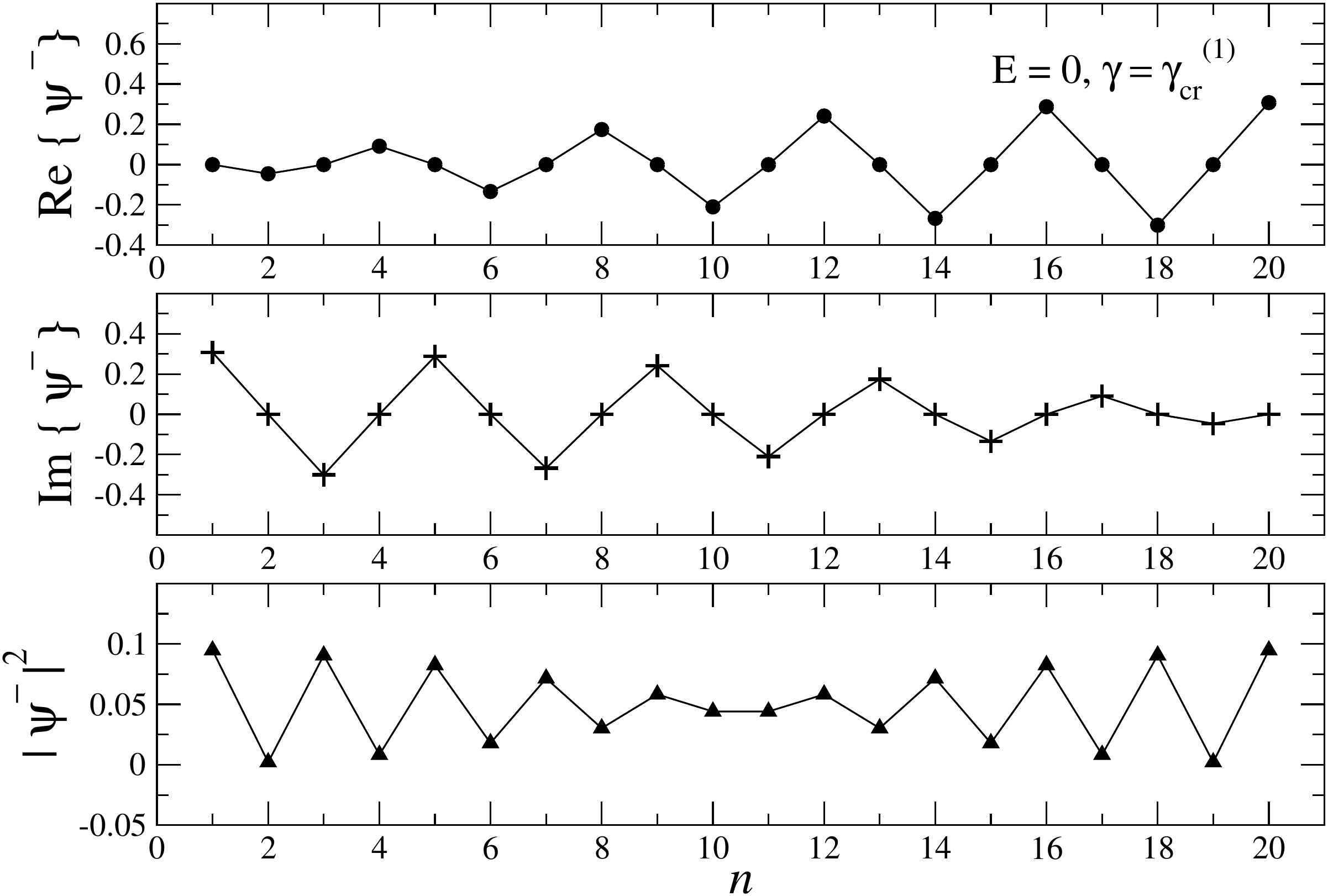}}
\caption{\footnotesize Degenerate eigenstates corresponding to
energy $E_s=0$, for a system with $\gamma=\gamma_{cr}^{(1)}$ and
$N=10$.} \label{eigens-0}
\end{figure}

\section{Unbounded model}

\subsection{Scattering states}

Below we consider the problem of transmission through the bounded model
with the PT-symmetric potential defined by Eq.(\ref{pot_pt}) for
$n=1,...,2N$, with $N$ standing for the total number of basic cells.
In contrast with the bounded model, now we assume that the model is
attached to perfect semi-infinite leads, specifically,
$\epsilon_{n}=0$ for $n>2N$ and $n\leq 0$. We also assume an
incoming plane wave from the right side of the system, therefore,
the left lead is occupied by the transmitted wave only. One can use
the transfer matrix approach which allows to find $\psi_{2N+1}$ and
$\psi_{2N}$ due to the following relation,
\begin{equation}
\left( \begin{array}{c}
\psi_{2N+1}  \\
\psi_{2N}  \\
\end{array} \right)=
      \mathcal{M}^N
\left(\begin{array}{c}
      \psi_{1}  \\
      \psi_{0}  \\
      \end{array}\right),\quad\quad\quad\mathcal{M}=
      \left(\begin{array}{cc}
      E-i\gamma & -1\\
      1         &  0\\
      \end{array}\right)
\left(\begin{array}{cc}
      E+i\gamma & -1\\
      1         &  0\\
      \end{array}\right).
\label{product-M}
\end{equation}
As one can see, the problem is reduced to the corresponding
dynamical system,
\begin{equation}
\left( \begin{array}{c}
\psi_{n+1}  \\
\psi_{n}  \\
\end{array} \right)=
\left(\begin{array}{cc}
      E-i\gamma & -1\\
      1         &  0\\
      \end{array}\right)
\left(\begin{array}{cc}
      E+i\gamma & -1\\
      1         &  0\\
      \end{array}\right)
\left(\begin{array}{c}
      \psi_{n-1}  \\
      \psi_{n-2}  \\
      \end{array}\right)=
      \mathcal{M}
\left(\begin{array}{c}
      \psi_{n-1}  \\
      \psi_{n-2}  \\
      \end{array}\right),
\end{equation}
for $``n"$ {\it even}. Note that in this representation the index
$n$ denoting the sites can be treated as the discrete "time". The
solution can be written in terms of eigenvectors
$\overrightarrow{\xi_1}, \overrightarrow{\xi_2}$ and eigenvalues
$\lambda_{1}, \lambda_{2}$ of the matrix $\mathcal{M}$,
\begin{equation}
 \left( \begin{array}{c}
\psi_{n+1}\\
\psi_{n}  \\
\end{array} \right)=
B\lambda_{1}^{n/2}\overrightarrow{\xi_1}+A\lambda_{2}^{n/2}\overrightarrow{\xi_2},
\label{psi-psi}
\end{equation}
with constants $A$ and $B$ determined by the initial conditions
\cite{book},
\begin{equation}
 \psi_0=1,\qquad\psi_1=e^{-ik}.
 \label{in_cond}
\end{equation}
One can show that the eigenvalues of matrix $\mathcal{M}$ are given
by,
\begin{equation}
 \lambda_{1,2}=\frac{E^2+\gamma^2}{2}-1\mp\frac{1}{2}\sqrt{(E^2+\gamma^2-2)^2-4},
\end{equation}
and the corresponding eigenvectors are
\begin{equation}
\overrightarrow{\xi_1}=
 \left( \begin{array}{c}
\frac{1+\lambda_1}{E+i\gamma}  \\
1  \\
\end{array} \right),\qquad
\overrightarrow{\xi_2}=
 \left( \begin{array}{c}
\frac{1+\lambda_2}{E+i\gamma}  \\
1  \\
\end{array} \right).
\end{equation}
Using the parametrization, similar to that in
Eqs.(\ref{spectrum},\ref{beta-n},\ref{2-delta}),
\begin{equation}
 E=2 \cos \mu\sin \beta, \quad\frac{\gamma}{2 \cos \mu}=\cos \beta, \quad
 \delta=-ie^{i\beta},
\end{equation}
the eigenvalues and eigenvectors can be written as follows,
\begin{equation}
 \lambda_{1,2}=e^{\mp2i\mu},
 \label{eig_re}
\end{equation}
and
\begin{equation}
\overrightarrow{\xi_1}=
 \left( \begin{array}{c}
\delta e^{-i\mu}  \\
1  \\
\end{array} \right),\qquad
\overrightarrow{\xi_2}=
 \left( \begin{array}{c}
\delta e^{i\mu}  \\
1  \\
\end{array} \right).
\label{ksi}
\end{equation}

Here it is important to stress the difference between the problem of
spectrum and eigenstates (for bounded model), and the scattering
problem (for the unbounded model). In the former we fix the parameter
$\gamma$ and boundary conditions, in order to obtain the
corresponding energy levels. In contrast, the (real) energy $E$ in
the scattering problem is the energy of the scattering wave,
therefore, it is a free parameter. The physical meaning of $\mu$ can
be compared with that of Bloch wave number that emerges in 1D
periodic structures with $N \rightarrow \infty$. In our model $N$
can get any value, nevertheless, the variable $\mu$ plays the role
of a wave number inside the sample. According to the dispersion
relation,
\begin{equation}
 4\cos^2\mu=E^2+\gamma^2,
 \label{def-mu}
\end{equation}
the value of $\mu$ can be either real or imaginary depending on
whether $E^2+\gamma^2\le 4$ or $E^2+\gamma^2>4$, respectively. We
have to note that in contrast with the bounded model the parameter
$\delta$ has one sign only due to specific conditions
(\ref{in_cond}) for $\psi_0$ and $\psi_1$ corresponding to the {\it iteration}
of $\psi_n$ from the left to right side of the sample. Note, however, that the {\it propagation} of the scattering wave occurs from the right to left size of the structure.

By inserting Eqs.(\ref{ksi},\ref{eig_re}) into Eq.(\ref{psi-psi}) we
get,
\begin{eqnarray}
\psi_{n+1}&=&\delta(Ae^{i(n+1)\mu}+Be^{-i(n+1)\mu}),\nonumber\\
\psi_{n}&=&Ae^{in\mu}+Be^{-in\mu}, \label{psi-final}
\end{eqnarray}
with $``n"$ {\it even}. These expressions are of the same form as
Eq.(\ref{psi-m}), however, they have different meaning. In contrast
with Eq.(\ref{psi-m}) defining the eigenstates of stationary
Schr\"{o}dinger equation, here $\psi_n$ are components of scattering
states inside the sample attached to the leads.

In order to determine the constants $A$ and $B$ one has to use the
initial conditions (\ref{in_cond}). As a result, we obtain,
\begin{equation}
 A=\frac{ie^{-i\beta}e^{-ik}-e^{-i\mu}}{2i\sin \mu},
 \quad B=1-\frac{ie^{-i\beta}e^{-ik}-e^{-i\mu}}{2i\sin \mu}.
 \label{A+B}
\end{equation}
Eqs.(\ref{psi-final},\ref{A+B}) fully determine the scattering
states for $\mu \neq 0$, for both real and imaginary value of
$\mu$. Fig.(\ref{scat-st}) shows the structure of the scattering
states in both cases. In the former case the on-site probabilities
inside the sample are periodic functions with respect to the index $n$. In contrast, when $\mu$ is
imaginary, there are two terms in the expression for on-site
probabilities, one of which is an exponentially increasing function
of $n$, and another is a decreasing function. Therefore, formally
one can speak about the localization of scattering states, and the
exponential decrease of $T$ for $E^2+\gamma^2>4$ can be directly
related to the localization length. Note that such a localization
occurs in the absence of disorder, the fact which has been noticed
when discussing the properties of scattering for the model with
constant gain or loss only \cite{GJ95,PMB96,MPB97,JJ97,NN98,JS99,JLS99,JJ00,H06,GCS12}.
\begin{figure}[ht]
\centering
\subfigure[]{\includegraphics[scale=0.3]{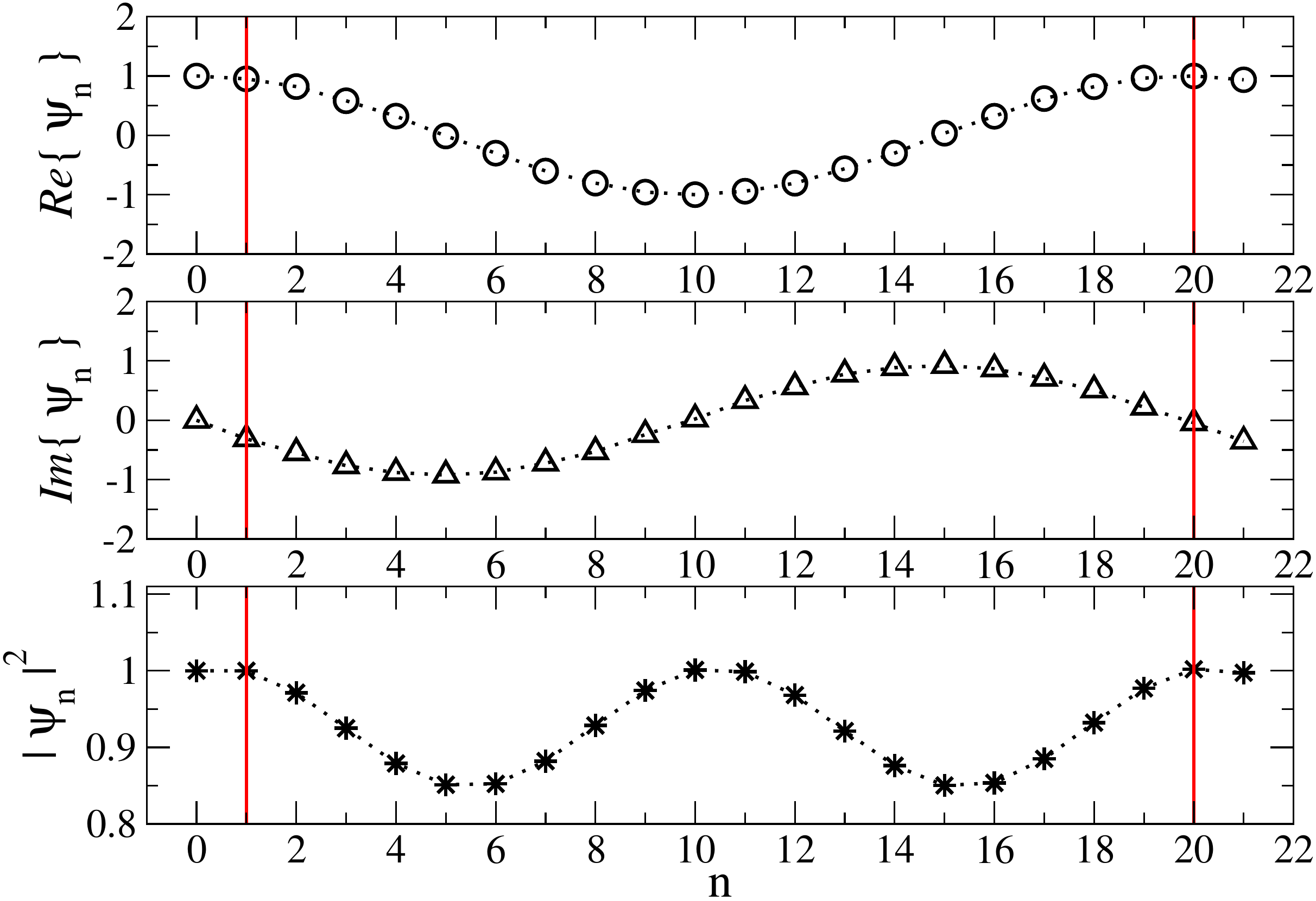}}
\subfigure[]{\includegraphics[scale=0.3]{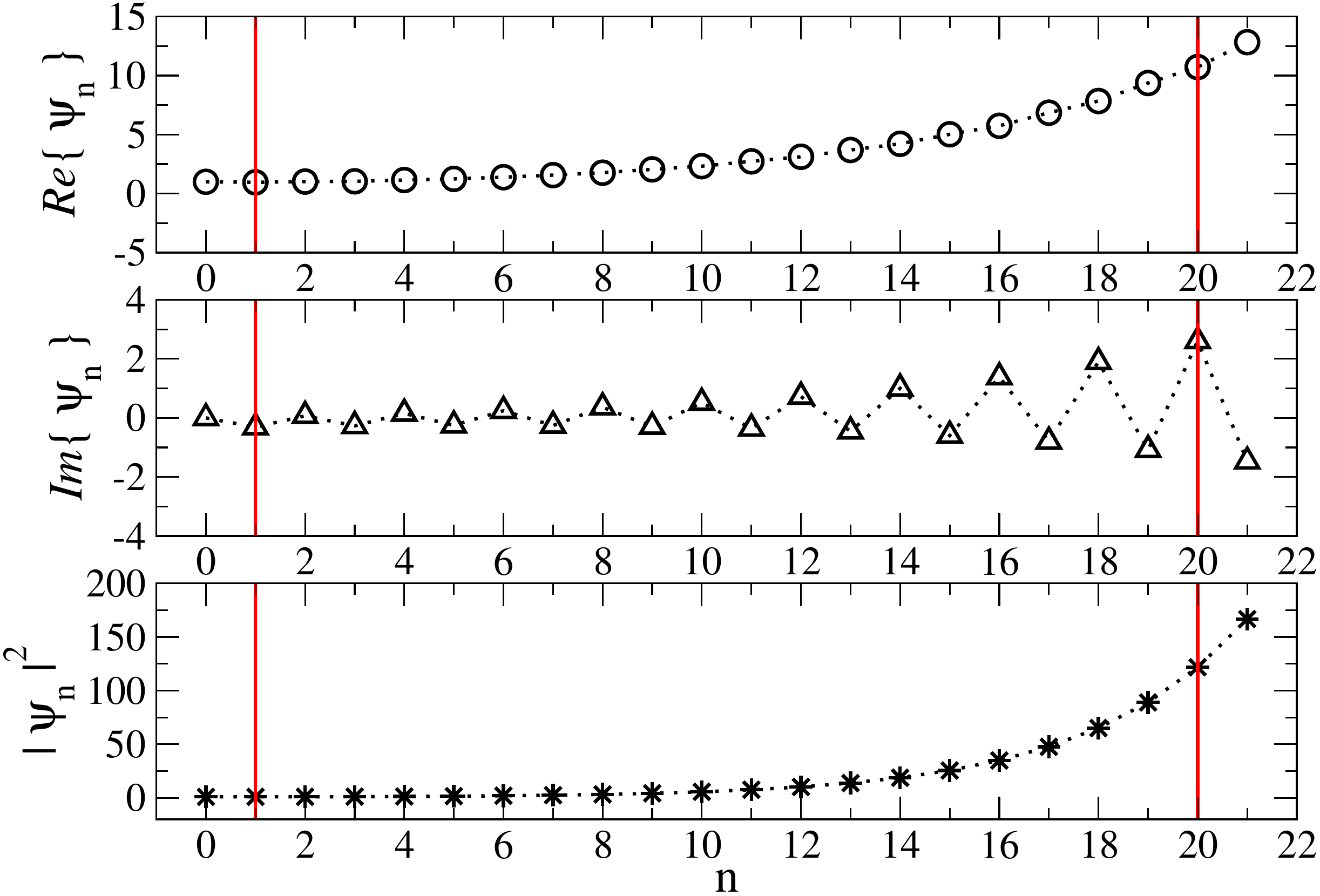}} \caption{
\footnotesize Scattering states for $E=1.9$ and $N=10$. (a)
Scattering state for $\gamma=0.05$ such that $E^2+\gamma^2<4$ and
$\mu$ real. (b) Scattering state for $\gamma=0.7$ such that
$E^2+\gamma^2>4$ and $\mu$ imaginary.} \label{scat-st}
\end{figure}

We can get an estimate of the rate of increase of $|\psi_{n}|^2$
with the use of Eqs.(\ref{psi-final},\ref{A+B}) applied to the case
when $E^2+\gamma^2>4$. By introducing $\mu=i\phi$ for $\phi>0$, it
is sufficient to analyze only the scattering state for even sites,
\begin{equation}
 |\psi_{n}|^2=e^{2n\phi}\left(\frac{C_1}{2}+C_2+1\right)+e^{-2n\phi}\left(\frac{C_1}{2}
 \right)-C_1-C_2,
 \label{inc-ss}
\end{equation}
where
\begin{equation}
 C_1=\frac{(\cosh\phi-\sin(\beta+k))e^{\phi}}{\sinh^2\phi},
 \quad C_2=\frac{\sin(\beta+k)-e^{\phi}}{\sinh\phi}.
\end{equation}
Taking $n>>1$, one gets,
\begin{equation}
 |\psi_{n}|^2\propto e^{2n\phi},
 \label{psi-increase}
\end{equation}
therefore, the rate of exponential increase is given by $2\phi$.
Correspondingly, the localization length can be defined as
\begin{equation}
 \frac{1}{\ell_{\infty}}=2\phi\approx\sqrt{E^2+\gamma^2-4}.
 \label{loc-length}
\end{equation}
Here the estimate is given for $\gamma \ll 1$. Since in this case
the energy $E$ has to be close to the band edge $E=\pm 2$, one can
write, $|E|=2 - \Delta$ where $\Delta$ is the distance from the band
edge. By assuming $\gamma^2 \gg 4 \Delta$, one can get the
simplified estimate for the localization length, $\ell_{\infty}
\approx 1/\gamma$.

For the special case with $E^2+\gamma^2=4$ we have $\mu=0$, and from
Eq.(\ref{psi-final}) one gets, $A=1$ and $B=0$. Therefore, the
on-site probabilities $|\psi_{n}|^2$ are all equal to unity as
Fig.\ref{scat-z} clearly manifests. As is expected, and will be shown rigorously below, the
transmission coefficient $T$ in this case equals 1. Since $\mu=0$,
this means that in this case the wave propagates through the
structure of size $N$ without change of its phase. Thus, such a
structure will be non-visible for an observer.
\begin{figure}[ht]
\centering
\includegraphics[scale=0.3]{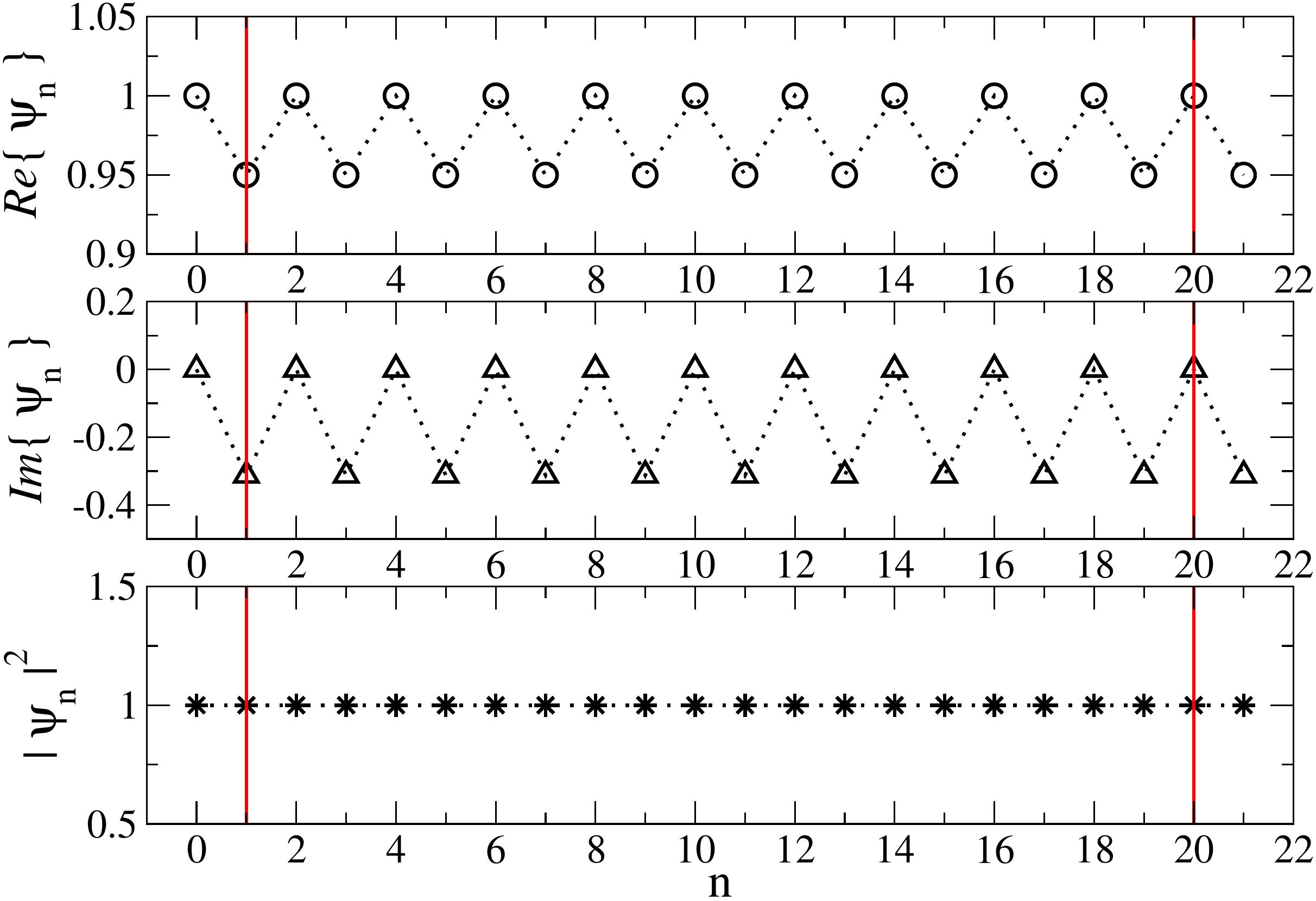}
\caption{ \footnotesize Scattering state for $E=1.9$, $N=10$ and
$\gamma=0.6245$ such that $E^2+\gamma^2=4$ and $\mu=
0$.} \label{scat-z}
\end{figure}
%
\subsection{Transmittance}

According to the general theory \cite{book}, the transmission
coefficient $T$ can be expressed as follows,
\begin{equation}
T=\frac{1}{|({\bf M}^{(N)})_{22}|^2}. \label{T-M}
\end{equation}
Here the matrix ${\bf M}^{(N)}$ emerges when the relation
(\ref{product-M}) is written in the representation of plane waves,
\begin{equation}
Q^{-1} \left( \begin{array}{c}
\psi_{N+1}  \\
\psi_{N}  \\
\end{array} \right)=
Q^{-1}\mathcal{M}^{(N)}QQ^{-1} \left(\begin{array}{c}
      \psi_{1}\\
      \psi_{0}\\
      \end{array}\right)=
{\bf M}^{(N)}Q^{-1} \left(\begin{array}{c}
      \psi_{1}\\
      \psi_{0}\\
      \end{array}\right),
      \label{8}
\end{equation}
where
\begin{equation}
Q=\left( \begin{array}{cc}
1 & 1  \\
e^{-ik} & e^{ik}  \\
\end{array} \right).
\label{9}
\end{equation}

After standard manipulations one can express the transmission
coefficient $T$ in terms of $\psi_{2N}$ and $\psi_{2N+1}$ (see
details in Ref.\cite{book}),
\begin{equation}
T=\frac{4\sin^2k}{|e^{-ik}\psi_{2N+1}-\psi_{2N}|^2}.
\label{T-from-M}
\end{equation}
In this expression $k$ is the wave number of incoming wave, and the
values of $\psi_{1}$ and $\psi_{0}$ have to be specified due to
Eq.(\ref{in_cond}). As is discussed above, the meaning of these
initial values is due to fixing the plane wave propagating to the
{\it left} from the sample (for $n\leq 1$), after an incoming wave
(electron) comes from the {\it right} of the sample (for $n>2N$).
Note that here the index $N$ corresponds to the cell consisting of
two cites with alternating gain and loss. Therefore, the total
number of cells in the scattering structure is $2N$.

Note that in contrast with the bounded model, here $E=2\cos k$ is the
energy of incoming plane wave, expressed through the wave vector
$k$. In order to obtain an expression for $T$ as a function of the
parameters $\gamma$, $E$ and $N$, we insert the expressions defined
by Eq.(\ref{psi-final}) into Eq.(\ref{T-from-M}), with the constants
$A$ and $B$ given by Eq.(\ref{A+B}). After some technical work, one can represent
the transmission coefficient in a quite compact form,
\begin{equation}
 T=\frac{1}{1-\frac{\gamma^2}{4\sin^2 k\cos^2\mu}\sin^2(2\mu N)},
 \label{T}
\end{equation}
This result is exact and valid for any values of control parameters,
$E$ and $\gamma$ for a fixed $N$. Let us discuss the main properties of the transmission with the use of this expression. First, one has to recall the relation for the
Bloch-like index $\mu$,
 \begin{equation}
 4 \cos^2 \mu=E^2+\gamma^2,\quad  \quad
  \mu=\left\{ \begin{array}{lll}
  \mbox {real \,\,\,\,\,\,\quad \quad for} \quad E^2+\gamma^2<4,\\
  \mbox {zero \,\,\,\,\,\,\quad \quad for} \quad E^2+\gamma^2=4,\\
  \mbox{imaginary \, for} \quad E^2+\gamma^2>4,
  \end{array}\right.
  \label{ran-pa}
 \end{equation}
with $2 \cos k =E$. One can see that when $\mu$ is real the
transmission $T$ is a {\it periodic} function of $N$. To the
contrary, when $\mu$ is imaginary, $\mu=i\phi$ ($\phi>0$), we have
$\cos \mu =\cosh \phi$, and $\sin(2\mu N)=i \sinh(2\phi N)$.
Therefore, $\ln T$ takes negative values resulting in its {\it
monotonic decrease} with an increase of $N$. From the theory of
disordered systems the exponential decrease of $T$ with the system
size $N$ is a specific property of the Anderson localization. Here,
we also can define the localization length with the use of the
standard relation,
\begin{equation}
 \frac{1}{\ell_{\infty}}=-\displaystyle \lim_{N \rightarrow \infty}\frac{\ln T}{2N},
 \label{def-ll}
\end{equation}
and treat the quantity $\ell_{\infty}$ as the localization length. In
the limit $N\rightarrow\infty$ one gets for $\ln T$,
\begin{equation}
\ln T\approx-4\phi N.
\end{equation}
Therefore, by inserting last expression into Eq.(\ref{def-ll}), we
arrive at the expression (\ref{loc-length}) for $\ell_{\infty}$
describing an exponential decrease of the scattering wave
propagating from the right to left of the structure.

As one can see from Eq.(\ref{T}), the transmission coefficient $T$
equals unity for $\mu=0$ which leads to the relation
$E^2+\gamma^2=4$ between the energy $E$ and the parameter $\gamma$.
For a fixed energy $E$ this relation defines the critical value
$\gamma_{cr}$,
\begin{equation}
E^2+\gamma^2_{cr}=4, \label{gam-cr}
\end{equation}
such that for $\gamma < \gamma_{cr}$ the parameter $\mu$ is real,
therefore, the dependence of $\ln T$ is periodic function of $N$,
see Fig.\ref{ln-T}. Note that in this case the transmission
coefficient is larger than 1, apart from the set of resonances with
$T=1$, defined by the relation $N\approx m\pi/2\mu$ with $m$ as positive
integer. The maximum value of $T$ observed in this regime is located
at the points $N\approx m\pi/4\mu$ with $m$ positive odd integers, leading
to the value $\ln T=-\ln(1-\frac{\gamma^2}{4\sin^2k\cos^2\mu})$. For
$\gamma > \gamma_{cr}$ the value of $\mu$ is imaginary, therefore,
the transmission $T$ decreasing with an increase of $N$, see
Fig.\ref{ln-T}. Note that for specific value $\gamma=\gamma_{cr}$
the logarithm of the transmission coefficient vanishes for {\it any}
$N$.
\begin{figure}[ht]
\centering
\includegraphics[scale=0.35]{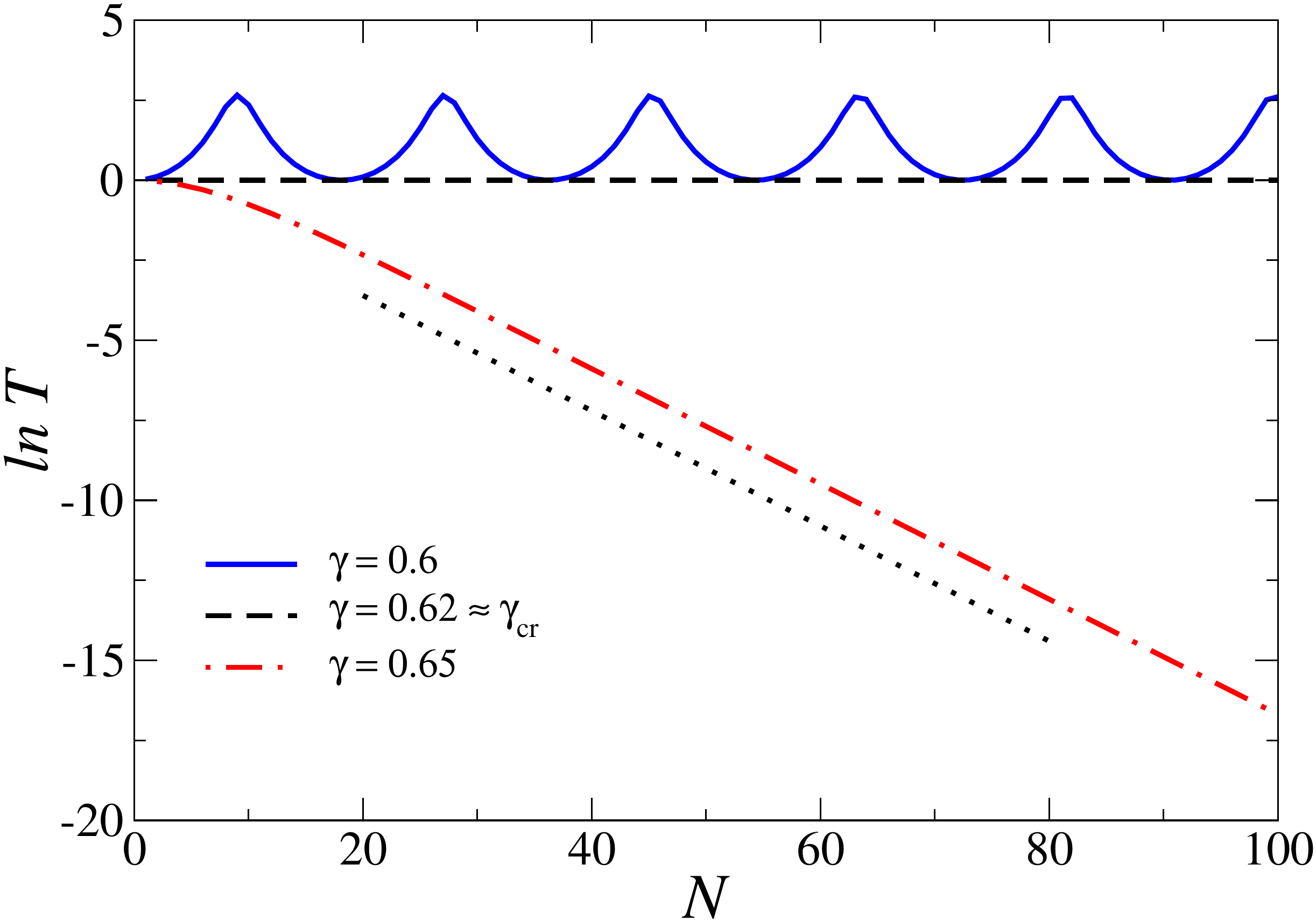}
\caption{\footnotesize (Color online) The $N-$dependence of $\ln T$ for
$\gamma=0.6,\, 0.62,\, 0.65$, and fixed energy $E=1.9$. When $\gamma
< \gamma_{cr}\approx 0.62$, the logarithm of $T$ is a positive
periodic function with the fundamental period $\pi/2\mu$. When
$\gamma\approx\gamma_{cr}$ we have $\ln T=0$. Finally, when $\gamma
>\gamma_{cr}$, the value of $\ln T$ is negative, with a monotonically decreasing
dependence. The dotted line shows the analytic estimate of $\ln T$
for $\gamma >\gamma_{cr}$.} \label{ln-T}
\end{figure}

The behavior of $\ln T$ as a function of $E$ can be seen in
Fig.\ref{lnvsE} for the case $N=2$ (therefore, with four cites).
Here we can see that as $E$ approaches the band edges the
transmission coefficient vanishes. One can detect the resonant energies $E_r$ for which $T=1$. In
general, they are given by the expression,
\begin{equation}
 E_r=\pm\sqrt{4\cos^2\left(\frac{m\pi}{2N}\right)-\gamma^2},\quad (m=0,1,\ldots,N).
 \label{res-E}
\end{equation}
Note that the number of such resonances in dependence on the energy
depends not only on $N$ but also on the parameter $\gamma$, because
$E_r$ is real only when
$4\cos^2\left(\frac{m\pi}{2N}\right)>\gamma^2$. Therefore, for a
sufficiently large $\gamma$ all resonances may disappear. Also we
would like to note that the first resonance emerging for $m=0$
corresponds to $\mu=0$, at this point $\ln T$ changes its sign.
\begin{figure}[ht]
\centering
\includegraphics[scale=0.35]{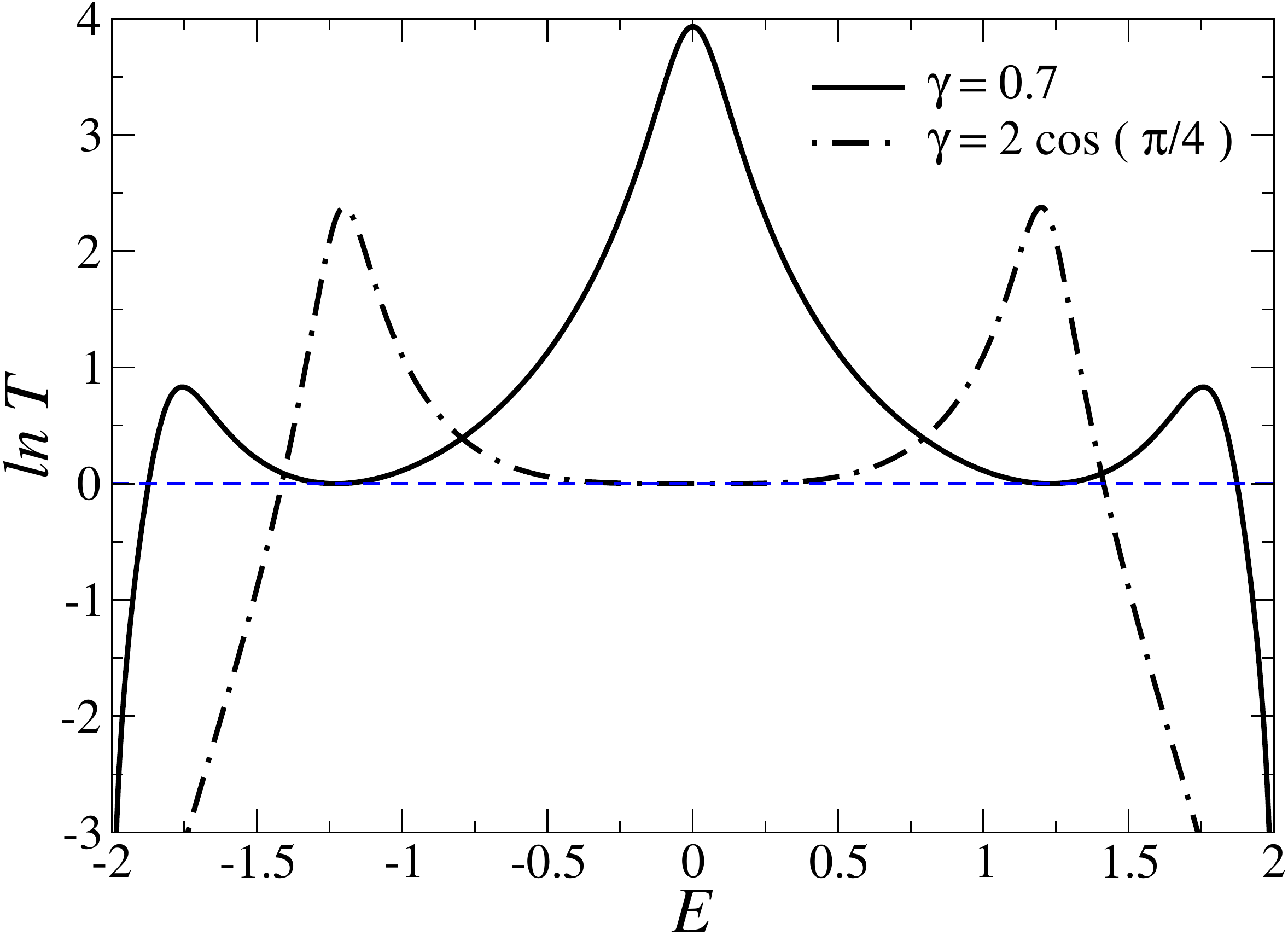}
\caption{\footnotesize Energy dependence of $\ln T$ for $N=2$ with
$\gamma=0.7$ and $\gamma=2\cos(\pi/4)$. Note that for the latter value of $\gamma$, $T\approx1$ in a quite
large region of energy around $E=0$.} \label{lnvsE}
\end{figure}

Finally, by Fig.\ref{lnT-gam} we demonstrate the behavior of $\ln T$
as a function of $\gamma$. Here again one can see the resonances
defined by the relation,
\begin{equation}
 \gamma_r=\sqrt{4\cos^2\left(\frac{m\pi}{2N}\right)-E^2},\quad
 (m=0,1,\ldots,N).
 \label{res-ga}
\end{equation}
The total number of these resonances is determined by the size $N$
of the sample and by the energy $E$. One can see that for $N=2$, the transmission for small $\gamma$ is close to unity
practically for any energy.

\begin{figure}[ht]
\centering
\subfigure[] {\includegraphics[scale=0.35]{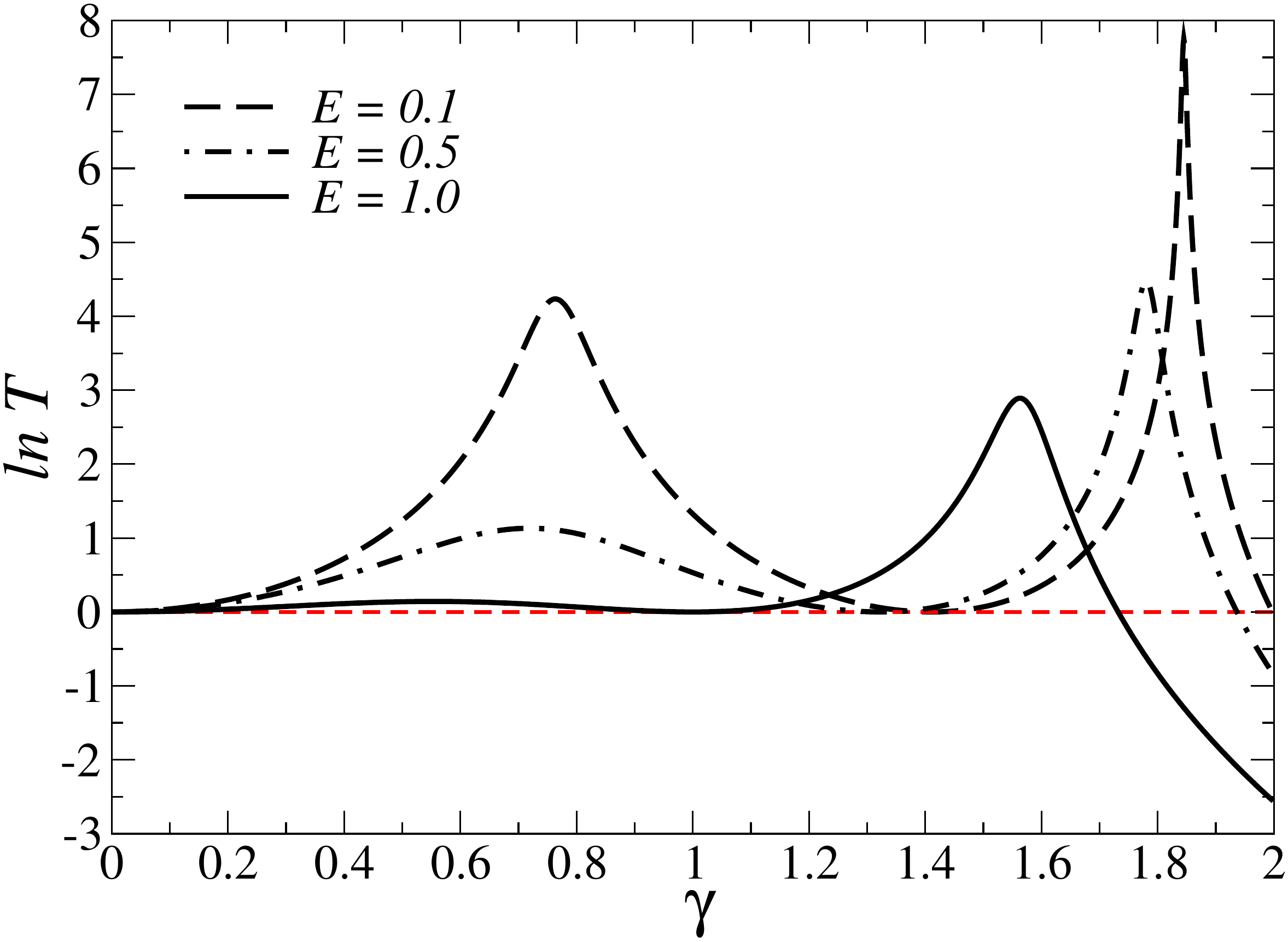}}
\caption{\footnotesize $\ln T$ versus $\gamma$ for $N=2$ and
$E=0.1,0.5,1.0$.} \label{lnT-gam}
\end{figure}

\subsection{Reflectance}

The analytical expressions for the reflectance can be obtained by
the transfer matrix (see Eq.(\ref{8})),
\begin{equation}
{\bf M}^{(N)}=Q^{-1}\mathcal{M}^{(N)}Q.
      \label{Tm}
\end{equation}
After some algebra one can express the elements of the transfer
matrix in terms of the parameters of our model,
\begin{eqnarray}
 ({\bf M}^{(N)})_{1,1}&=&\frac{1}{\sin k\sin\mu}
 \left(\cos(2\mu N)\sin\mu\sin k-i\sin(2\mu N)(\cos\mu\cos k-\sin\beta)\right),\\
 ({\bf M}^{(N)})_{1,2}&=&\frac{1}{\sin k\sin\mu}
 \left(ie^{ik}\sin(2\mu N)(\sin(\beta+k)-\cos\mu)\right),\\
 ({\bf M}^{(N)})_{2,1}&=&\frac{1}{\sin k\sin\mu}
 \left(ie^{-ik}\sin(2\mu N)(\cos\mu-\sin(\beta-k))\right),\\
 ({\bf M}^{(N)})_{2,2}&=&\frac{1}{\sin k\sin\mu}
 \left(\cos(2\mu N)\sin\mu\sin k-i\sin(2\mu N)(\sin\beta-\cos\mu\cos k)\right).
 \label{Tme}
\end{eqnarray}
According to the definition, the right and left reflectances, $R_R$
and $R_L$, are defined as follows,
\begin{equation}
 R_R=\left|\frac{({\bf M}^{(N)})_{1,2}}{({\bf M}^{(N)})_{2,2}}\right|^2,
 \quad R_L=\left|\frac{({\bf M}^{(N)})_{2,1}}{({\bf
 M}^{(N)})_{2,2}}\right|^2.
 \label{Ref}
\end{equation}
Therefore, the reflectance for the wave incident from the right side
of the system, one gets,
\begin{equation}
 R_R=\frac{(\cos\mu-\sin(\beta+k))^2}{(\sin\beta-\cos\mu\cos k)^2+(\cot(2\mu N)\sin\mu\sin k)^2},
 \label{RR}
\end{equation}
and for the wave incident from the left side we have,
\begin{equation}
 R_L=\frac{(\cos\mu-\sin(\beta-k))^2}{(\sin\beta-\cos\mu\cos k)^2+(\cot(2\mu N)\sin\mu\sin k)^2},
 \label{RL}
\end{equation}
For PT-symmetric systems, the generalized relation between the
transmittance and reflectances, reads \cite{GCS12},
\begin{equation}
 \sqrt{R_RR_L}=|1-T|,
 \label{gen-ur}
\end{equation}
which is fulfilled for our model. Here $T$ is the left or right
transmittance of the system given by, the definition, $T=1/|({\bf
M}^{(N)})_{2,2}|^2$, see Eq.(\ref{T-M}). As a check we see that
when $\gamma\rightarrow0$, both reflectances $R_{R,L}\rightarrow0$
which is the expected result.

Let us now discuss the properties of $R_R$ and $R_L$ when $\gamma$
is less, equal or larger than $\gamma_{cr}$ which is defined by the
relation, $E^2+\gamma_{cr}^2=4$. The data in
Figs.\ref{rr_vsN},\ref{rl_vsN} show the behavior of both
reflectances for the above three cases, as a function of $N$. For
$\gamma<\gamma_{cr}$ both reflectances are periodic functions of $N$
with the same resonant points at $N=m\pi/2\mu$, with $m$ as a positive integer. Of course, due to
the relation (\ref{gen-ur}) these points are equal to those when
$T=1$. For $\gamma=\gamma_{cr}$ one can see a very different
behavior of both reflectances: while $R_R\approx0$, the left
reflectance $R_L$ grows with $N$. In order to analyze this behavior
one can use Eqs.(\ref{RR}-\ref{RL}). Note that these equations are
not valid for the case when $\gamma$ is exactly equal to
$\gamma_{cr}$, nevertheless they give the correct limit.
Specifically, taking $\mu\rightarrow0$ we get $R_R\approx 0$ and
$R_L\propto N^2$. This effect was termed {\it unidirectional
reflectivity} in Ref.\cite{24} where the authors have studied the
PT symmetric model with finite width of barriers. When
$\gamma>\gamma_{cr}$ one can observe that both reflectances approach
a constant value quite rapidly, the fact that can be easily deduced
from Eqs.(\ref{RR}-\ref{RL}).
\begin{figure}[ht]
\centering
\includegraphics[scale=0.3]{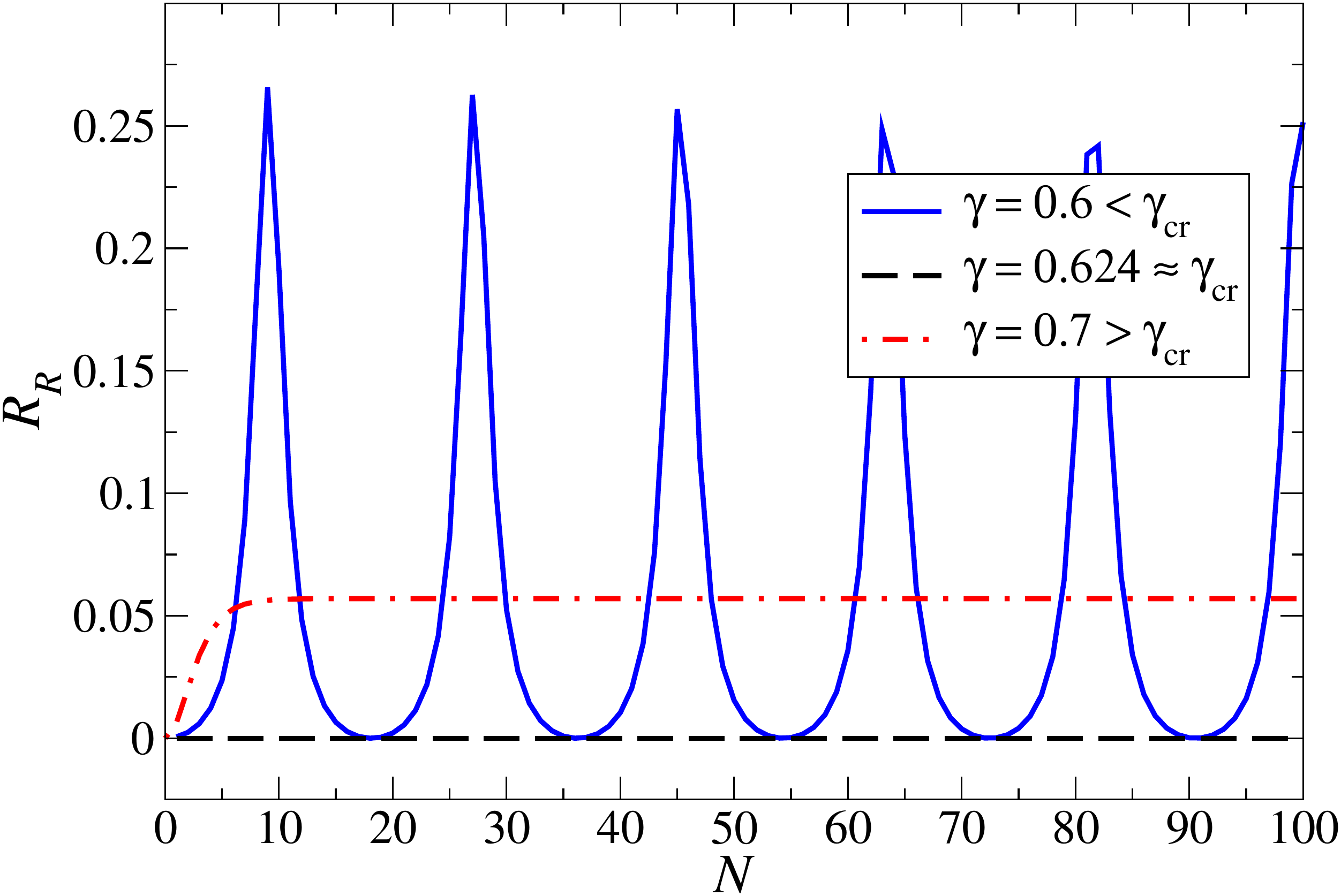}
\caption{\footnotesize (Color online) Right reflectance versus $N$ for $E=1.9$
and $\gamma=0.6, 0.624, 0.7$, with $\gamma_{cr}\approx0.624$.}
\label{rr_vsN}
\end{figure}
\begin{figure}[ht]
\centering
\subfigure[]{\includegraphics[scale=0.3]{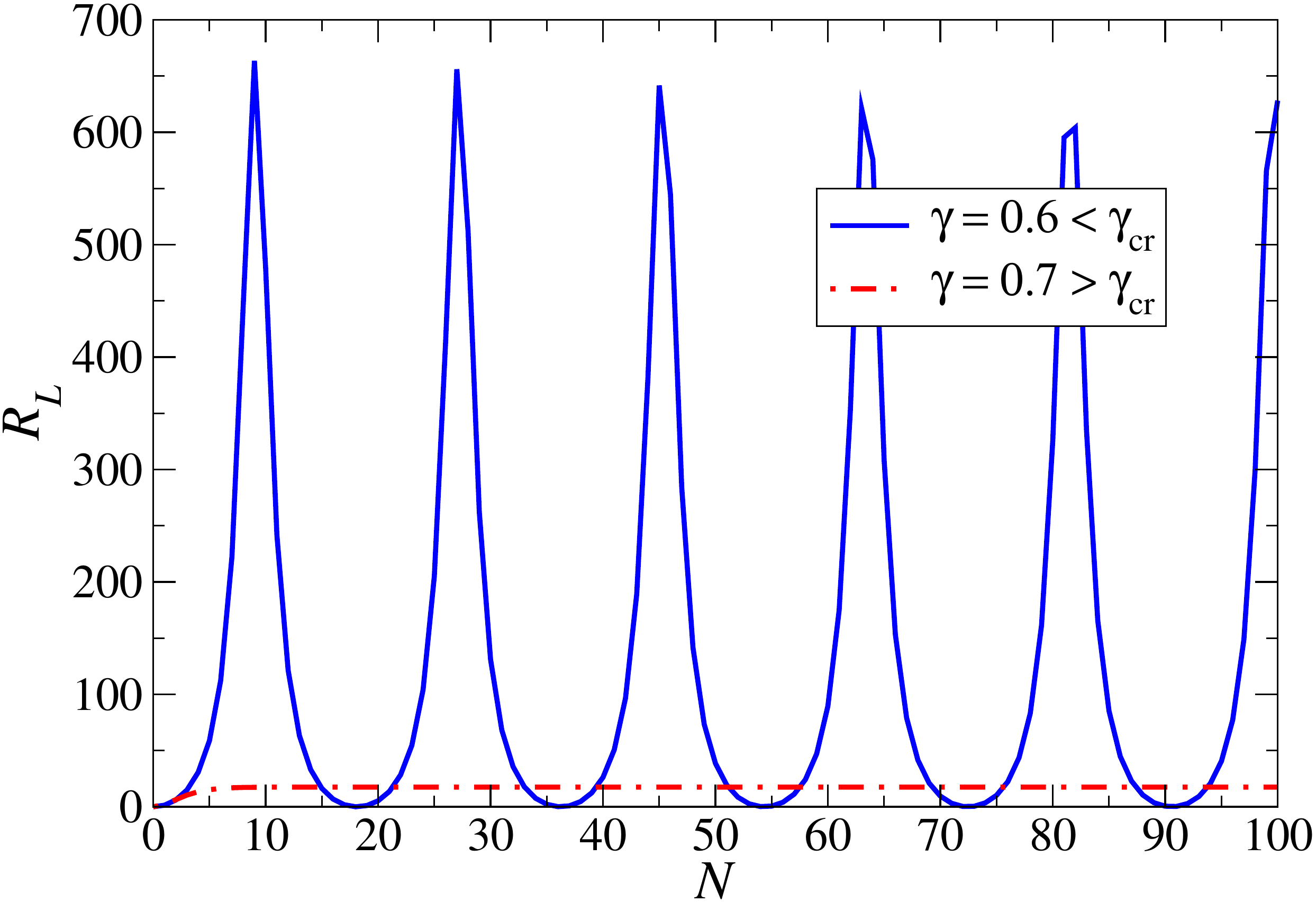}}
\subfigure[]{\includegraphics[scale=0.3]{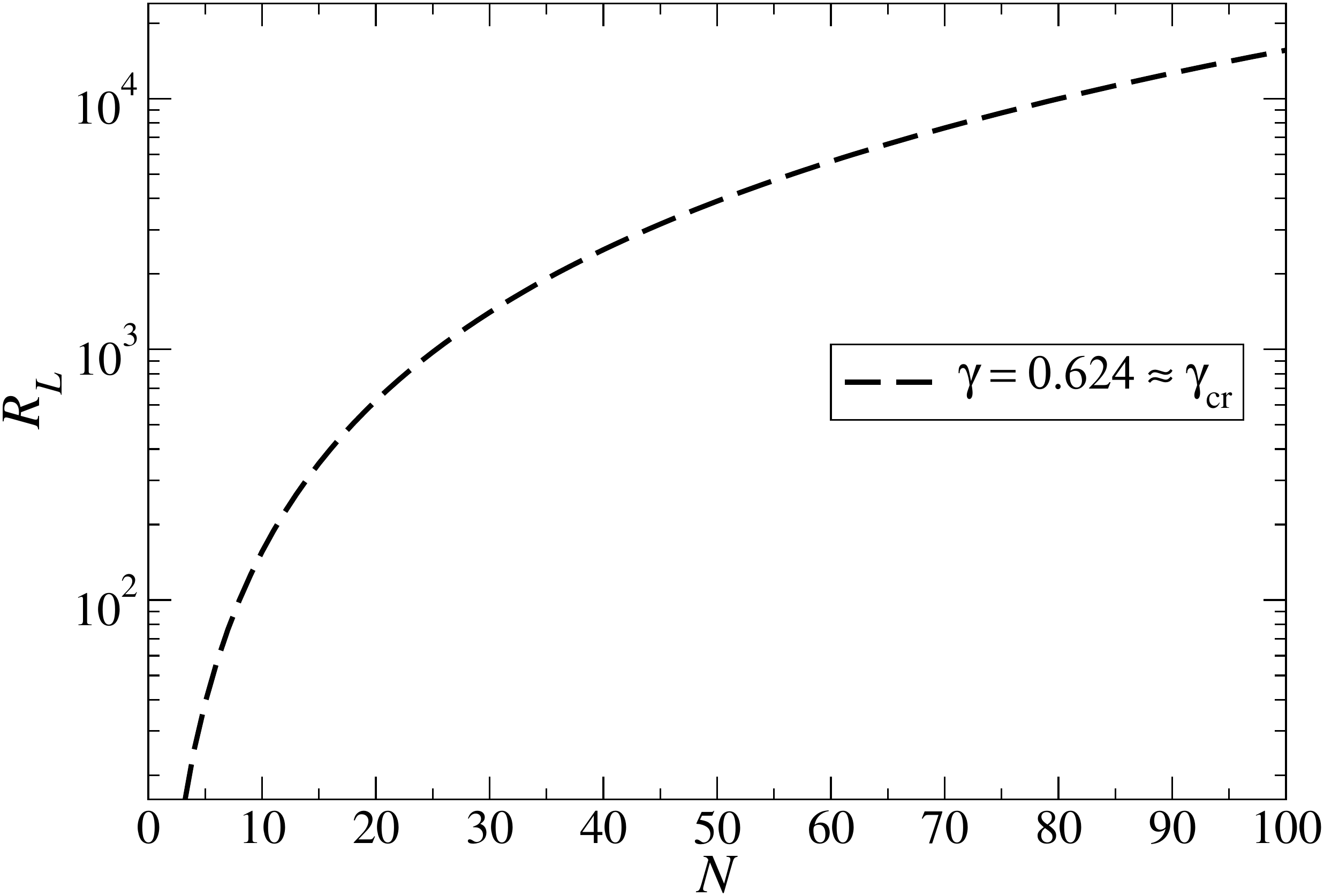}}
\caption{\footnotesize (Color online) Left reflectance versus $N$ for $E=1.9$.
(a) $\gamma=0.6 < \gamma_{cr}$ and $\gamma = 0.7 > \gamma_{cr}$. (b)
$\gamma=0.624\approx\gamma_{cr}$.} \label{rl_vsN}
\end{figure}

Fig.\ref{rrlvsE} shows the dependence of the reflectances with
respect to $E$. From these data as well as from
Eqs.(\ref{RR},\ref{RL}), one can see that when $|E|\rightarrow 2$,
both reflectances, $R_{R,L}\rightarrow 1$. Due to the fact that the
dependence on $E$ is different for both reflectances they show a
different number of points where they vanish. Indeed, while for
$R_R$ the vanishing points are the same that the ones corresponding
to the resonant values of $T$, see Eq.(\ref{res-E}), for $R_L$ this
holds only when $E^2+\gamma^2<4$. The resonant value of $T$ for
$E^2+\gamma^2=4$ corresponds to an increase in $R_L$, not to a local
maximum, and to vanishing of $R_R$. This is due to the fact that at
this point we have $\cos\mu-\sin(\beta+k)=0$ and
$\cos\mu-\sin(\beta-k)=2\sin^2k$.
\begin{figure}[ht]
\vspace{1cm}
\centering \subfigure[]{\includegraphics[scale=0.3]{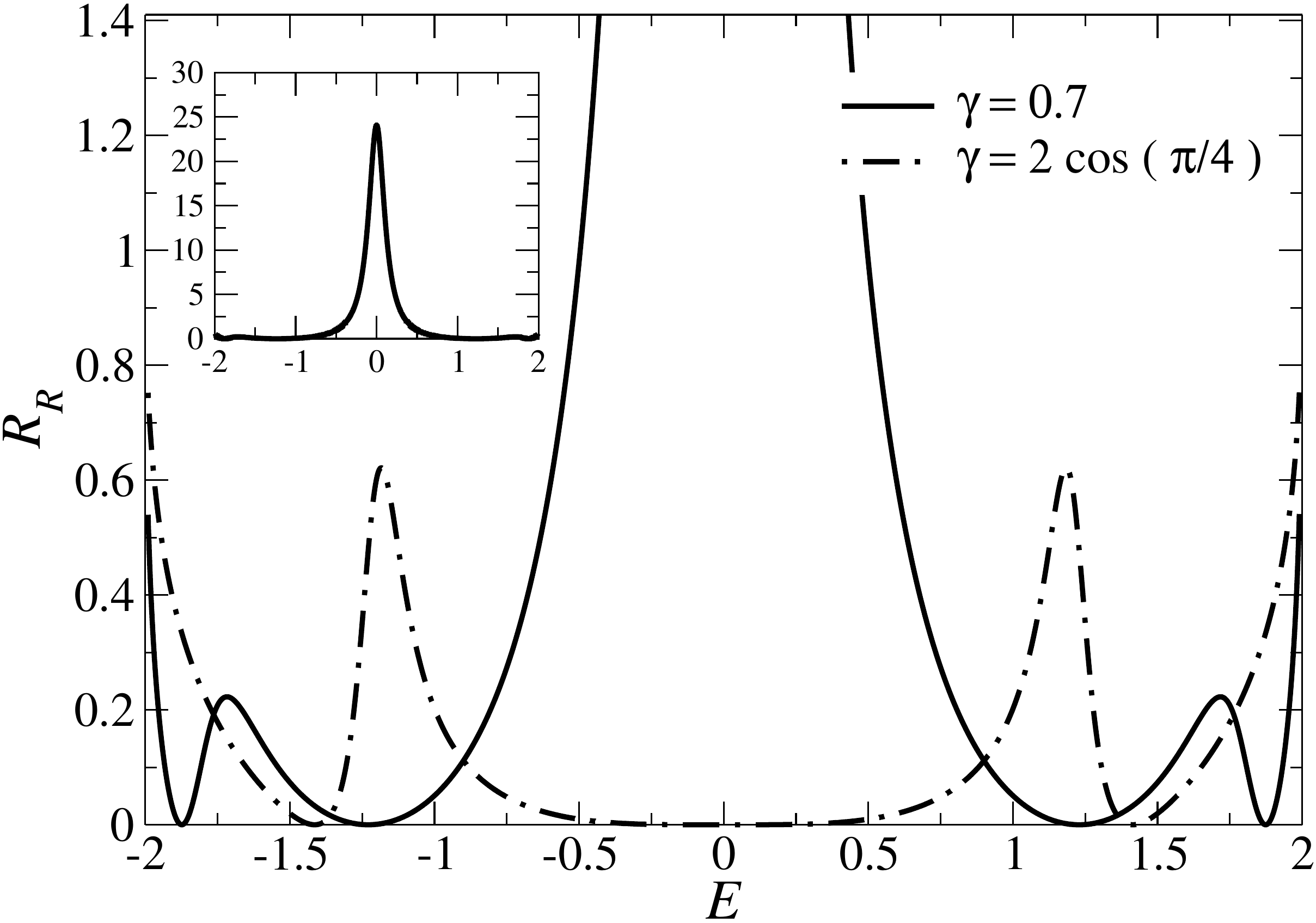}}
\subfigure[]{\includegraphics[scale=0.3]{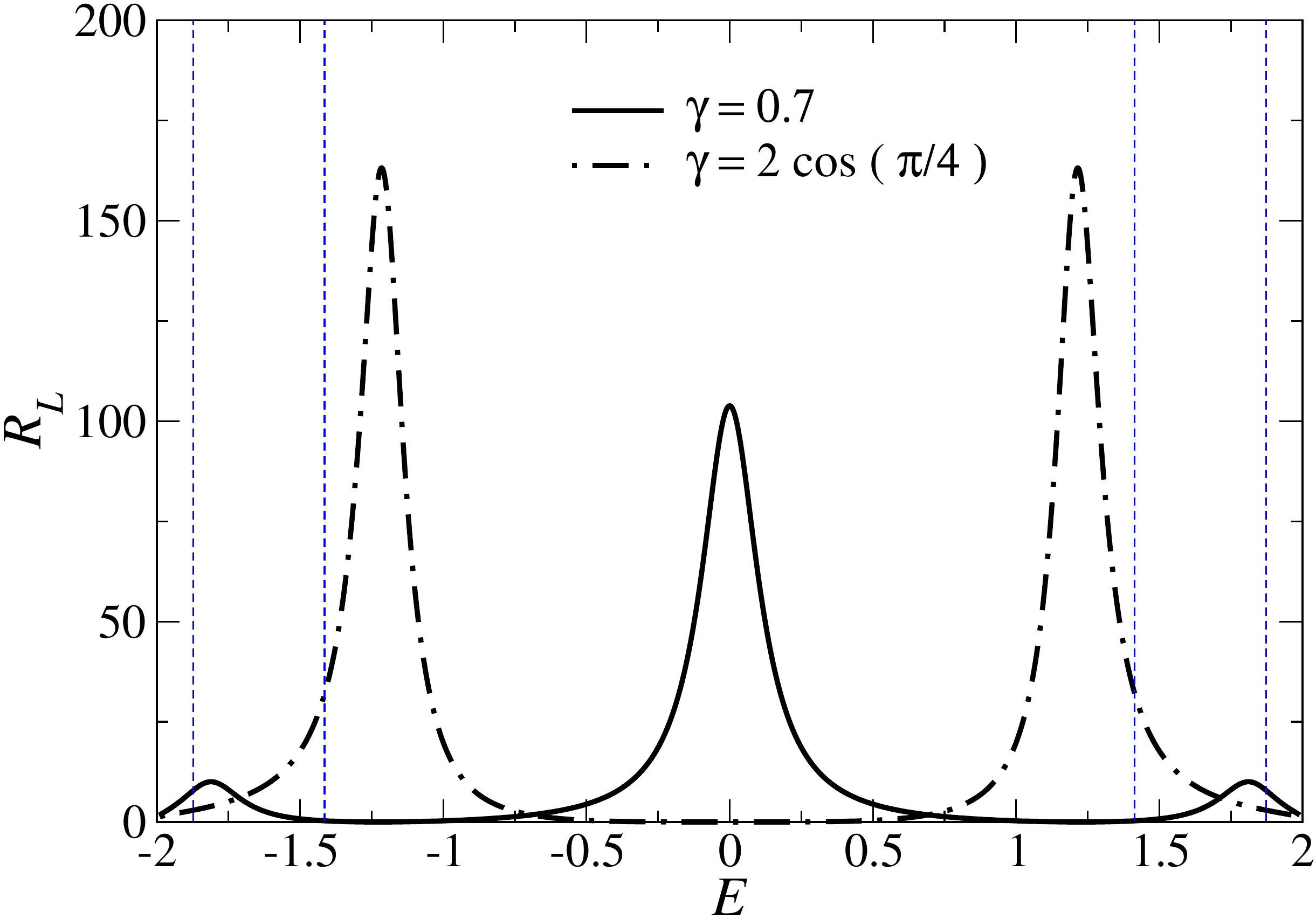}}
\caption{\footnotesize (a) Right reflectance versus $E$ for
$\gamma=0.7$ and $\gamma=2\cos(\pi/4)$ where the latter corresponds
to the resonant value of $T$ at the band center, $N=2$. Inset shows
full plot for $\gamma=2\cos(\pi/4)$. (b) Left reflectance versus $E$
for the same parameters as in (a). Here vertical lines mark the
points where $E^2+\gamma^2=4$.} \label{rrlvsE}
\end{figure}

\subsection{Discussion}

The approach which we have used in this paper allows us to relate
the properties of spectra and eigenstates for ``bounded" model to the
structure of scattering states for ``unbounded" model, as well as to the
properties of transmission and reflection for finite samples
attached to perfect leads. The key ingredient for the relation
between the two models is the expression (\ref{main-1}) determining
the energy spectrum in the bounded model. This expression has to be
compared with Eq.(\ref{def-mu}) which defines the parameter $\mu$
emerging in the unbounded model. One can see that both expressions have
the same structure, however, they have different meaning. In the
first case (bounded model) the energies $E_s$ are determined by the
parameter $\gamma$ and by the wave vector $k_s$ emerging due to the
zero or periodic boundary conditions. To the contrary, in the unbounded
model the energy $E$ is a free parameter corresponding to the energy
of an incident wave in the scattering problem, and the parameter
$\mu$ emerges in place of $k_s$. The physical meaning of this
parameter $\mu$ can be associated with the Bloch index for periodic
structures in the limit $N \rightarrow \infty$. However, our results
are also valid for {\it any} finite value of $N$, including $N=1$
(the model with only two sites with alternating gain and loss) and
$N=2$ (the model with four sites, see some results above). Thus, for
finite $N$ the parameter $\mu$ could be treated as the
generalization of the Bloch index.

The physical meaning of the parameter $\mu$, however, remans clear
even for finite $N$. As we have shown, it determines the structure
of the scattering states. Specifically, if $\mu$ is real, the
scattering states are extended inside the sample of size $N$, and
$\mu$ can be treated as the wave vector. A completely different
situation occurs for the imaginary values of $\mu$, for which the
scattering states consist of two components, one is exponentially
increasing function of the site index $n$ and another is
exponentially decreasing function of $n$. This fact is extremely
important in view of the value of the transmission coefficient $T$.
It turns out that the specific boundary conditions corresponding to
the scattering problem (see Eq.(\ref{in_cond})) result in the
asymmetry for exponentially increasing/decreasing components in the
scattering states. This fact directly leads to the exponential
decrease of $T$ in the limit of large $N$. As a result, one can
formally introduce the localization length $\ell_{\infty}$ in the
same way as it is done in the theory of disordered models, see
Eq.(\ref{loc-length}). In Section IV we have shown that the same value
of localization length can be obtained by analyzing the structure of
scattering states.

It should be stressed that the emergence of localization in the unbounded
model in the absence of disorder is due to the presence of gain and
loss only. Similar result is already known for the Anderson model
with vanishing disorder, with the loss or gain only \cite{GJ95,PMB96,MPB97,JJ97,NN98,JS99,JLS99,JJ00,H06,GCS12}.
The mathematical origin for this kind of localization is due to the
property of transfer matrices that determine all transport
characteristics of the scattering. To be more precise, one should
indicate that two eigenvalues $\lambda_{1,2}$ of the matrix
$\mathcal{M} $, see Eq.(\ref{product-M}), corresponding to the
periodic cell with two sites gain/loss, are related to each other
due to the simple relation, $\lambda_1 \lambda_2 = 1$. This means that
in the absence of gain and/or loss, the two eigenvalues are
complex with modulo one which corresponds to real values of $\mu$. On the other
hand, and this is the key point, in the presence of gain and/or loss
the two eigenvalues are real with $\lambda_1=1/\lambda_2$. This means
that in the structure of scattering states there are two components,
one is exponentially growing with $n$ and the other decreases exponentially
with $n$. This effect is quite unexpected since naively
one can expect that the presence of gain or loss leads to either
increase or decrease of scattering states versus $n$. However,
the actual result is due to the presence of two increasing/decreasing
components. The surprising fact is that in spite of these two
components in the structure of scattering states, the boundary
conditions {\it at one side} of a sample (specifically, at two
sites) effectively select one component which prevails in the limit $N
\rightarrow \infty$, see Eq.(\ref{psi-increase}).

From our analysis it is clear that the value of the localization
length, if it is defined for large enough $N$ in the formal way
(either from the structure of scattering states or due to the
logarithm $\ln T $ of the transmission coefficient), can be obtained
simply from the expression for the Bloch-like index $\mu$. As we
have shown, it can be done due to the relation,
\begin{equation}
\frac{1}{\ell_{\infty}}=2 |\mu|, \quad \cosh^2 |\mu|  = \frac{1}{4}
(E^2 +\gamma^2).
 \label{general}
\end{equation}
for $\mu$ imaginary. The generalization of this relation is
straightforward for any kind of model allowing to be expressed by
the transfer matrix describing {\it one} periodic cell. In our case, the
basic cell consists of two sites, however, in general
it could consist of any number of sites. Then, having the expression
for $\mu$ one can easily obtain the localization length
$\ell_{\infty}$ from the first relation in Eq.(\ref{general}). The
same approach can be used for more complicated periodic cells in the
model, for example, with different amount $\gamma_1$ of gain and
$\gamma_2$ of loss. Note that the presence of the PT-symmetry is not
important for such an analysis. As is now clear, the PT-symmetry
allows to have a quite specific situation with real eigenvalues of
energy for the eigenstates which is equivalent to real values of the
Bloch-like parameter $\mu$ for scattering states.

One of the central results of our study is a quite simple expression for
the transmission coefficient $T$, see Eq.(\ref{T}). It is rigorously
obtained for {\it any} value of $\gamma$ and $N$ and any value of
energy $E$ of an incident wave. This expression allows one to
analyze various situations emerging in the scattering. It shows that
for real value of $\mu$ for which $\gamma < \gamma_{cr}$ the
transmission coefficient $T$ is larger than 1, apart from specific
values of $N$ for which $T=1$. On the contrary, for imaginary $\mu$
corresponding to $\gamma > \gamma_{cr}$, the value of $T$ is less
than 1. This is related to the onset of localization of scattering
states as is shown in Section IV-A. At the exceptional point $\mu=0$
with $\gamma=\gamma_{cr}$, the transmission is perfect, $T=1$, and,
together with the vanishing change of phase for the scattering
states, this leads to the invisibility of the structure \cite{M13}.
Although this effect emerges only for specific values of energy $E$,
from Fig.\ref{lnvsE} one can see that the transmission coefficient is very close to unity
in a quite large region of $E$.

As for the reflectance, our results show an emergence of
``unidirectional reflectivity" observed in Ref.\cite{24} for the
PT symmetric model with a finite width of barriers. This effect
is more pronounced for large values of $N$. The analytical
expressions give a full information about the difference between
left and right reflectances in dependence on the model parameters.

In conclusion, our analytical results reveal the mechanism for the
emergence of the localized scattering states which is entirely due
to the presence of the gain/loss terms in the non-Hermitian
Hamiltonian. It should be stressed, that the onset of such a
localization is not related to the PT symmetry, it emerges due to
the imaginary value of the Bloch-like parameter $\mu$. This
conclusion is quite general and can occur in any non-Hermitian model, thus allowing to express the localization in terms of this parameter $\mu$ only. The role of the PT symmetry
is just to have three regimes in the same model, governed by the
value of $\mu$ (or, the same, by the relation between $E$
and $\gamma$). These three regimes are: the transmission with $T>1$ due to extended scattering
states, perfect transmission for $\mu=0$ and any $N$ (resulting
in the invisibility), and the regime with $T<1$ emerging due to localized
scattering states.
\section{acknowledgments}

F.M.I. is thankful to L.I.Deych, A.A.Lisyansky, N.N.Makarov, V.V.Sokolov, V.G.Zelevinsky for fruitful discussions; he also acknowledges the support from CONACyT grants N-161665 and 133375.

\end{document}